\documentclass[galaxies,review,accept,pdftex,oneauthor]{Definitions/mdpi} 

\firstpage{1} 
\pubvolume{10}
\issuenum{10}
\articlenumber{75}
\pubyear{2022}
\copyrightyear{2022}
\externaleditor{Academic Editors: Jaziel Goulart Coelho and Rita C. Anjos}
\datereceived{30 May 2022} 
\dateaccepted{11 June 2022} 
\datepublished{17 June 2022} 
\hreflink{https://doi.org/10.3390/
galaxies10030075} 

\Title{Determination of the 
	Cosmic-Ray Chemical Composition: Open Issues and Prospects}

\TitleCitation{Determination of the Cosmic-Ray Chemical Composition: Open Issues and Prospects}

\Author{Alberto Daniel Supanitsky \orcidA{}}

\AuthorNames{A. D. Supanitsky}

\AuthorCitation{Supanitsky, A.D.}

\address[1]{%
Instituto de Tecnolog\'ias en Detecci\'on y Astropart\'iculas 
(CNEA, CONICET, UNSAM), Centro At\'omico Constituyentes, San Mart\'in, Buenos Aires CP B1650KNA, Argentina; daniel.supanitsky@iteda.cnea.gov.ar} 

\abstract{Cosmic rays are relativistic particles that come to the Earth from outer space. Despite 
a great effort made in both experimental and theoretical research, their origin is still unknown. 
One of the main keys to understand their nature is the determination of its chemical composition 
as a function of primary energy. In this paper, we review the measurements of the mass composition 
above $10^{15}$ eV\@. We first summarize the main aspects of air shower physics that are relevant 
in composition analyses. We discuss the composition measurements made by using optical, radio, and 
surface detectors and the limitations imposed by current high-energy hadronic interaction models 
that are used to interpret the experimental data. We also review the photons and neutrinos searches
conducted in different experiments, which, in addition to being important to understand the nature of 
cosmic rays, can provide relevant information related to the abundance of heavy or light elements 
in the flux at the highest energies. Finally, we summarize the future composition measurements that 
are currently being planned or under development.}


\begin{document}

\section{Introduction}

Despite having been discovered more than a century ago, the origin and nature of cosmic rays remain 
uncertain. However, in the last decades, great progress has been made in the understanding of this 
phenomenon. The cosmic-ray energy spectrum extends from $\sim$$10^9$ eV up to more than $10^{20}$ eV\@.
Due to the very fast drop of the flux for increasing values of primary energy, the most energetic 
particles cannot be detected directly. Therefore, for energies above $\sim$$10^{15}$ eV, the cosmic 
rays are detected by measuring the extensive air showers (EASs) that they generate due to their 
interaction with the air molecules. Ground-based detectors that cover very large areas are built to 
measure those EASs. Although these observatories allow to detect very small values of the cosmic-ray 
flux, the characteristics of the primary particle have to be reconstructed from the EAS data, which 
makes the data analyses more complex, increasing the sources of systematic uncertainties.

There are two main different types of detectors used to detect the cosmic rays of energies above 
$10^{15}$ eV\@. These are: surface detectors that measure the secondary particles of the EAS at 
ground level, and radiation detectors that measure the electromagnetic radiation emitted during the 
shower development in the atmosphere. Due to the steep drop of the flux, the observatories dedicated 
to detect cosmic rays of higher energies require larger collection areas.
 
\begin{figure}[ht!]
\centering
\includegraphics[width=12cm]{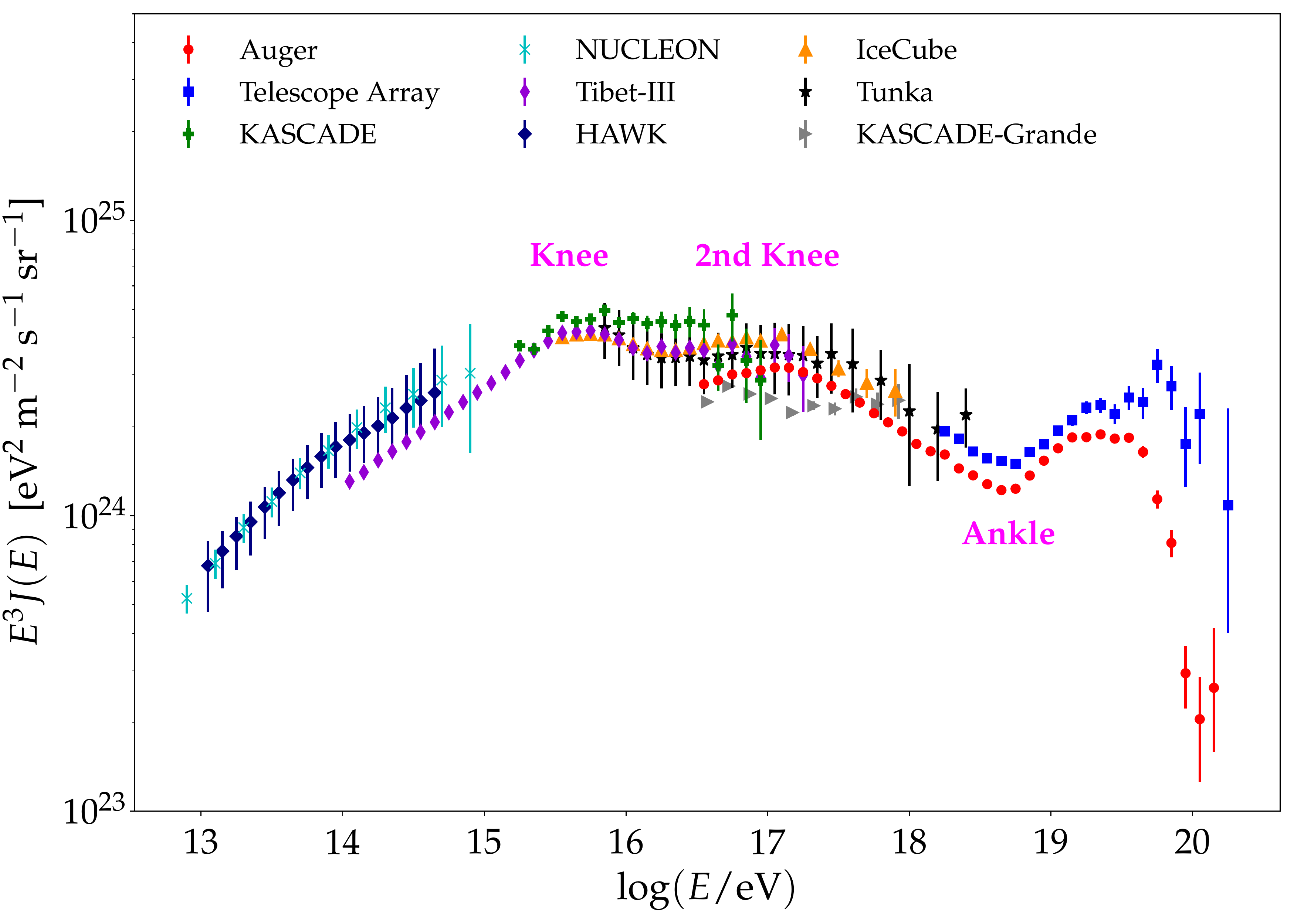}
\caption{Cosmic-ray flux, multiplied by $E^3$, as a function of the logarithmic energy. The data are
taken from Refs.~\cite{AugerJ:19,TAJ:19,KJ:11,NUCJ:19,TibetJ:08,HawkJ:17,IceCubeJ:19,TunkaJ:20,KGJ:15}. 
Adapted from~\cite{Carmeloevoli:21}. \label{CRFlux}}
\end{figure}

There are three main observables used to study the cosmic rays: the energy spectrum, the chemical 
composition of the primary particles, and the distribution of their arrival directions. The energy 
spectrum of the cosmic rays has been measured by several experiments as can be seen in Figure 
\ref{CRFlux}. It can be described as a broken power law with spectral indexes close to $\gamma=3$. 
It presents four main features: ($i$) the knee, a steepening of the flux located at $\sim$$10^{15.6}$ eV, 
($ii$) the second knee, a second steepening of the flux located at $\sim$$10^{17}$ eV, ($iii$) the 
ankle, a hardening of the flux located at $\sim$$10^{18.7}$ eV, and ($iv$) the suppression located 
at $\sim$$10^{19.6}$ eV\@. This last feature is a strong steepening of the flux with $\gamma>4$; it 
may be an exponential drop of the flux but the current statistics are too low to determine it.

It has long been believed that the galactic cosmic rays dominate the flux at low energies and extragalactic 
ones at high energies. This is based on the known inefficiencies of the galactic sources to accelerate 
particles to the highest energies. Moreover, the Pierre Auger Observatory (hereafter Auger) has found 
strong evidence about the extragalactic origin of the cosmic rays with energies above $10^{18.9}$ eV 
\cite{AugerSc:17}. The transition between the galactic and extragalactic components is still an open 
problem in cosmic-ray physics. In fact, the scenarios in which the transition takes place at the 
ankle are disfavored by the Auger data~\cite{Auger:12}. Therefore, the transition between these two 
components should take place somewhere between $\sim$$10^{16}$ and $\sim$$10^{18}$ eV\@.          
     
The determination of the mass composition as a function of the primary energy is crucial to understand 
many aspects of cosmic rays. However, the main limitation is that the composition determination is 
based on the comparison of experimental data with EAS simulations. Because the high-energy hadronic 
interactions relevant for the cosmic rays are unknown, models that extrapolate lower-energy accelerator 
data are used in shower simulations. This practice introduces systematic uncertainties that can be quite 
large depending on the shower observable under consideration.

The relevant parameter for the motion of charged particles in magnetic fields is the rigidity, 
$R= E/(Z \times e)$, where $E$ is the energy of the particle, $Z$ the charge number, and $e$ the absolute
value of the electron charge. The trajectories of particles with smaller rigidity values are strongly 
deviated by a given magnetic field. The acceleration of the cosmic rays and also its propagation on 
the galactic and extragalactic magnetic field depends on the rigidity. Therefore, the composition 
information is relevant for the study of the cosmic-ray sources and its propagation through the Galaxy 
and the intergalactic medium. In fact, due to the intensity of the galactic magnetic field, the 
identification of extragalactic sources is possible only considering light cosmic rays at the highest 
energies (large rigidity values)~\cite{Harari:14}.       

The composition information is also very important for the determination of the transition between 
the galactic and extragalactic components and the origin of the ankle~\cite{Kampert:12,Aloisio:12}.
The composition studies are also relevant to constrain the models of the suppression~\cite{AugerPrime:16}.
The composition is also key information to study the high-energy hadronic interactions beyond the reach 
of man-made accelerators~\cite{AugerPAir:12,TAPAir:20,Lipari:20}. 

The determination of the composition beyond the ankle is also important to predict the fluxes of cosmogenic
photons and neutrinos. These particles are generated as a by-product of the propagation of high-energy 
nuclei in the intergalactic medium~\cite{Gaisser:16}. In particular, they are generated due to the interaction 
of the cosmic rays with the radiation field present in the universe. Higher neutrino and photon fluxes are 
expected for lighter composition profiles at the highest energies (see, for instance, Ref.~\cite{Kampert:12}). 
Moreover, the measurement of the cosmogenic neutrino and photon fluxes can be very useful to constrain the 
composition in the suppression region~\cite{Hooper:11,Vliet:19}.

This paper is organized as follows: In Section~\ref{EAS}, we discuss the main aspects of the physics of the 
EAS oriented to composition analyses. In Section~\ref{CompoMeas}, we review the composition measurements performed
above $10^{15}$ eV by using optical, radio, and surface detectors. In Section~\ref{PhNu}, we review the photons
and neutrinos searches at the highest energies conducted in several experiments. In Section~\ref{Future}, we discuss 
the future of the composition measurements with current and future detectors. Finally, the main conclusions are 
discussed in Section~\ref{Conc}.

\section{Physics of Extensive Air Showers}
\label{EAS}

The characteristics of the EASs initiated by the cosmic rays depend on the primary type. The differences
among the EASs initiated by different primary types are used to identify the nature of the primary. Even 
though the detailed knowledge of the shower characteristics is studied through full Monte Carlo simulations,
there are simple models that allow to understand the main physical properties of the EAS. A simple model of
electromagnetic showers, i.e.,~initiated by photons, electrons, or positrons, was first introduced by 
Heitler~\cite{Heitler:44} and was extended to hadronic showers by Matthews~\cite{Matthews:05}.     
       
In the Heitler model, two processes are responsible for the electromagnetic shower development: Bremsstrahlung 
and pair production undergone by high-energy photons and electrons or positrons, respectively. Each electromagnetic
particle (photons, electrons, and positrons) undergoes a two-body splitting at a fixed distance 
$d=\lambda_\text{r} \ln 2$, where $\lambda_\text{r}$ is the radiation length in the medium (for the EAS, the medium 
is the atmosphere). The energy of each mother particle is distributed equally between the two daughter particles. 
The creation of particles ceases when its energy becomes smaller than the one necessary for undergoing Bremsstrahlung 
or pair production. Heitler takes this critical energy, $\mathcal{E}_\text{c}^e$, as the energy for which the 
radiative energy loss is equal to the collisional energy loss. In this simple model, the maximum of the shower is 
reached when the energy of each electromagnetic particle is equal to the critical energy. Therefore, 
$E/2^{n_\text{c}} = \mathcal{E}_\text{c}^e$, where $E$ is the primary energy and $n_\text{c}$ is the number of 
splitting lengths required to reach the maximum. The mean atmospheric depth at which the shower maximum is reached 
is given by $\langle X_{\textrm{max}}^{\textrm{em}} \rangle = n_\text{c} \lambda_\text{r} \ln 2$, which, after writing 
$n_\text{c}$ as a function of the primary energy and the critical energy, becomes
\begin{equation}
\langle X_{\textrm{max}}^{\textrm{em}} \rangle = \lambda_\text{r} \ln\left( \frac{E}{\mathcal{E}_\text{c}^e} \right).
\label{XmaxEM}
\end{equation}              
This simple model predicts quite well the values of $\langle X_{\textrm{max}}^{\textrm{em}} \rangle$ obtained 
from simulations. Note that $\langle X_{\textrm{max}}^{\textrm{em}} \rangle$ is linearly dependent on the 
logarithmic energy. 

In the extension of the Heitler model for the hadronic showers developed by Matthews (here on referred to as the 
Heitler--Matthews model), the atmosphere is also divided by layers of fixed thickness. Both the primary particle 
and each secondary hadron interact after passing through one layer of the atmosphere. Each hadron produces 
$N_{\text{ch}}$ charged pions ($\pi^{\pm}$) and $N_{\text{ch}}/2$ neutral pions ($\pi^{0}$). It is assumed that 
the neutral pions decay into two photons immediately after being created. These high-energy photons initiate 
electromagnetic showers. The charged pions continue their multiplication process until they all reach the critical 
energy, $\mathcal{E}_\text{c}^\pi$, which is taken as the energy for which the pion decay length is equal to the 
thickness of one layer. At this point, all charged pions decay producing muons and antimuons. Therefore, the hadronic 
showers are composed of three main components: electromagnetic, muonic, and hadronic (neutrinos and antineutrinos 
are also created in EASs, mainly in charged pion decays).

Let us consider first showers initiated by protons. Assuming that the energy is equally distributed among all 
created particles, the number of layers required to reach the point in which all charged pions decay is given by
\begin{equation}
n_\text{c} = \frac{\ln\left( E/\mathcal{E}_\text{c}^\pi \right)}{\ln\left( \frac{3}{2} N_{\text{ch}} \right)},
\label{nc}
\end{equation}              
where $E$ is the energy of the primary proton. As mentioned before, the number of muons (hereafter muons refers 
to muons and antimuons) is equal to the number of charged pions that reached the critical energy, 
$N_\mu^{p} = N_{ch}^{n_c}$. Therefore, using Equation~(\ref{nc}), the following expression for the average number of muons 
is obtained
\begin{equation}
\langle N_\mu^{\text{p}} \rangle = \left( \frac{E}{\mathcal{E}_\text{c}^\pi} \right)^\beta,
\label{Nmu}
\end{equation}              
where 
\begin{equation}
\beta = \frac{\ln N_{\text{ch}}}{\ln\left( \frac{3}{2} N_{\text{ch}} \right)}.
\label{Beta}
\end{equation}
Therefore, the Heitler--Matthews model predicts a power-law dependence with the primary energy of the muons number in 
proton-initiated air showers.   

In the Heitler--Matthews model, it is assumed that, in the first interaction, one-third of the proton energy is injected 
into the electromagnetic channel. The neutral pions generated in the subsequent steps of the cascade also feed the 
electromagnetic channel, but the energy directed to this channel is smaller than the one corresponding to the first 
interaction. Therefore, in this model, only the first interaction is considered to calculate the mean value of 
$X_{\textrm{max}}$.

In the first interaction $N_{\text{ch}}/2$, neutral pions are generated. Then, the number of photons generated in 
the decay of the neutral pions is $N_{\text{ch}}$. The energy of each neutral pion generated in the first interaction 
is $E_{\pi^0\! , \, \text{f}} = E/(3 N_{\text{ch}}/2)$, and then the energy of each photon generated after the decays 
of the neutral pion is  $E_{\gamma,\, \text{f}} = E_{\pi^0\! , \, \text{f}}/2 = E/(3 N_{\text{ch}})$. Considering that 
the mean atmospheric depth of the first interaction is $X_0$ and that at this point $N_{\text{ch}}$ showers are initiated 
by photons of energy $E_{\gamma,\, \text{f}}$, the mean depth of the shower maximum is obtained
\begin{equation}
\langle X_{\textrm{max}}^{\text{p}} \rangle = X_0 + \lambda_\text{r} \ln\left( \frac{E}{3 N_{\text{ch}}\, %
\mathcal{E}_\text{c}^e} \right).
\label{Xmaxp}
\end{equation}

Equation (\ref{Xmaxp}) can be written in terms of the mean value of $X_{\textrm{max}}$ corresponding to 
photon-initiated showers. By using Equation~(\ref{XmaxEM}), the following expression is obtained
\begin{equation}
\langle X_{\textrm{max}}^{\text{p}} \rangle = \langle X_{\textrm{max}}^{\gamma} \rangle + X_0 - %
\lambda_\text{r} \ln\left( 3 N_{\text{ch}} \right),
\label{XmaxpEM}
\end{equation}
where $X_{\textrm{max}}^{\gamma}$ is the atmospheric depth of the maximum shower development of photon-initiated
showers. 

The elongation rate measures the rate of change of the mean value of $X_{\textrm{max}}$ with logarithmic 
energy. It is defined as 
\begin{equation}
D = \frac{d \langle X_{\textrm{max}} \rangle}{d \ln E}.
\label{ELRate}
\end{equation}
Note that if $D_{10} = d \langle X_{\textrm{max}} \rangle / d \log E$, then $D_{10} = \ln(10) D$.

The elongation rate of proton showers can be obtained from Equations (\ref{XmaxpEM}) and (\ref{ELRate}).
It is given by
\begin{equation}
D_\text{p} = D_\gamma +\frac{d}{d \ln E} \left( X_0 - \lambda_\text{r} \ln( 3 N_{\text{ch}}) \right),
\label{ELRateP}
\end{equation}
which implies that $D_\text{p} < D_\gamma$, because $N_{\text{ch}}$ increases and $X_0$ decreases with primary 
energy. This result is known as the Linsley's elongation rate theorem~\cite{Linsley:77}, which states that the 
elongation rate of showers initiated by any hadron cannot exceed the one corresponding to photon showers.    

The number of muons and the depth of the shower maximum for heavier primaries can be obtained from the superposition
model. In this simple model, a primary nucleus of mass number $A$ and energy $E$ is considered as $A$ independent nucleus
of energy $E/A$ each. This approximation relies on the fact that the binding energy of the nucleus is much smaller than
the primary energy (see Ref.~\cite{Kampert:12} for a more detailed discussion). Therefore, by using the superposition 
model, the following expressions are obtained
\begin{eqnarray}
&& \langle N_\mu^{A} \rangle = A\, \left( \frac{E}{\mathcal{E}_\text{c}^\pi A} \right)^\beta, \\
&& \langle X_{\textrm{max}}^{A} \rangle = X_0 + \lambda_\text{r} \ln\left( \frac{E}{3 N_{\text{ch}}\, %
\mathcal{E}_\text{c}^e \, A} \right),
\end{eqnarray}   
which, from Equations~(\ref{Nmu}) and (\ref{Xmaxp}), take the following form
\begin{eqnarray}
\label{NmuANmuPr}
&& \langle N_\mu^{A} \rangle = A^{1-\beta}\, \langle N_\mu^{\text{p}} \rangle, \\
\label{XmaxAXmaxPr}
&&\langle X_{\textrm{max}}^{A} \rangle = \langle X_{\textrm{max}}^{\text{p}} \rangle -\lambda_\text{r} \ln A.
\end{eqnarray}   

As mentioned before, the detailed study of the EAS characteristics is conducted from full Monte Carlo simulations. 
The Monte Carlo programs most used in the literature are CORSIKA~\cite{corsika} and AIRES~\cite{aires}.
Moreover, the CONEX~\cite{conex} program is widely used in the literature as it combines the Monte Carlo technique 
with the numerical solution of the cascade equations to obtain the longitudinal development of the showers, which 
considerably reduces  the computing time compared to the full Monte Carlo simulations. The main limitation of the 
EAS simulations originates from the lack of knowledge of the hadronic interactions at the highest energies. Models 
that extrapolate the low-energy accelerator data to the required energies are included in the Monte Carlo programs. 
The high-energy hadronic interaction models commonly used in the literature have been updated recently with the Large 
Hadron Collider (LHC) data. However, some discrepancies among the predictions obtained from different models 
are still present, even at LHC energies, due to the limited collider data in the central region, which is relevant 
for the EAS simulation. These differences in the shower observables predicted by different high-energy hadronic 
interaction models introduce large systematic uncertainties in composition analyses. The post-LHC high-energy 
hadronic interaction models most used in the literature are: EPOS-LHC~\cite{epos}, Sibyll 2.3d~\cite{sibyll}, and 
QGSJet-II.04~\cite{qgsjetII}.        

Despite the strong simplifications in the Heitler--Matthews model, it predicts quite well the functional 
dependence of the mean number of muons and the mean depth of the shower maximum with primary energy and mass number.
However, the mean value of $X_{\textrm{max}}^{A}$ predicted by the model is about 100 g cm$^{-2}$ smaller
than the one obtained from EAS simulations~\cite{Matthews:05}. In Ref.~\cite{Matthews:05}, it is suggested that 
these discrepancies arise due to the non-inclusion of the photons, originated in neutral pion decays, produced 
beyond the ones corresponding to the first interaction. In Ref.~\cite{Montanus:14}, it is found that this is not
the origin of the discrepancies. Moreover, in that work, it is proposed that the discrepancies arise from the 
assumption of a homogeneous distribution of energy among the particles created in each step of the cascade.  

As mentioned before, the Heitler--Matthews model predicts the functional dependence of 
$\langle X_{\textrm{max}}^{A} \rangle$ with $A$ and $E$ (see Equation~(\ref{XmaxAXmaxPr})). This is shown in the analysis
performed in Ref.~\cite{AugerXmax:13}, made based on EAS simulations, where it is found that the mean value of 
$X_{\textrm{max}}^{A}$ can be written as
\begin{equation}
\label{XmaxASims}
\langle X_{\textrm{max}}^{A} \rangle = \langle X_{\textrm{max}}^{\text{p}} \rangle + F_E \ln A,
\end{equation}
where $F_E$ is a function of primary energy. 

Regarding the number of muons, it is also found that its dependence with $A$ and $E$ can be approximated by the 
one predicted by the Heitler--Matthews model (see Equation~(\ref{NmuANmuPr})). Moreover, it can be seen that the 
$\beta$ parameter ranges from $0.85$ to $0.93$~\cite{Matthews:05}, such that the values close to $\beta = 0.93$ 
correspond to post-LHC models~\cite{Cazon:18}.  

From the previous discussion, it can be seen that the EAS generated by light nuclei develop deeper in the atmosphere
and have a lower muonic content than an EAS generated by heavy nuclei. Therefore, the $X_{\textrm{max}}$ parameter and
also any parameter closely related to the muon content of the showers (such as the number of muons at ground, measured 
at a given distance to the shower axis or even in a given distance range) can be used to determine the nature of the 
primary particle. Moreover, these two parameters showed to be the best to discriminate between proton and iron 
primaries (see, for instance, Ref.~\cite{Supanitsky:08}). 

The longitudinal development of the EAS is characterized by the number of charged particles present at a given slant 
depth, $N_{\text{ch}}(X)$. However, a quantity that is more closely related to the observation of the longitudinal
development of an EAS with fluorescence telescopes is the energy deposition rate $\frac{dE}{dX}$, i.e., the longitudinal 
profile, whose integral gives the total energy deposited in the atmosphere. Because $N_{\text{ch}}(X)$ is, to a very 
good approximation, proportional to the longitudinal profile, the maximum reached by both functions is located at 
approximately the same atmospheric depth.
     
\begin{figure}[ht!]
\centering
\includegraphics[width=6.5cm]{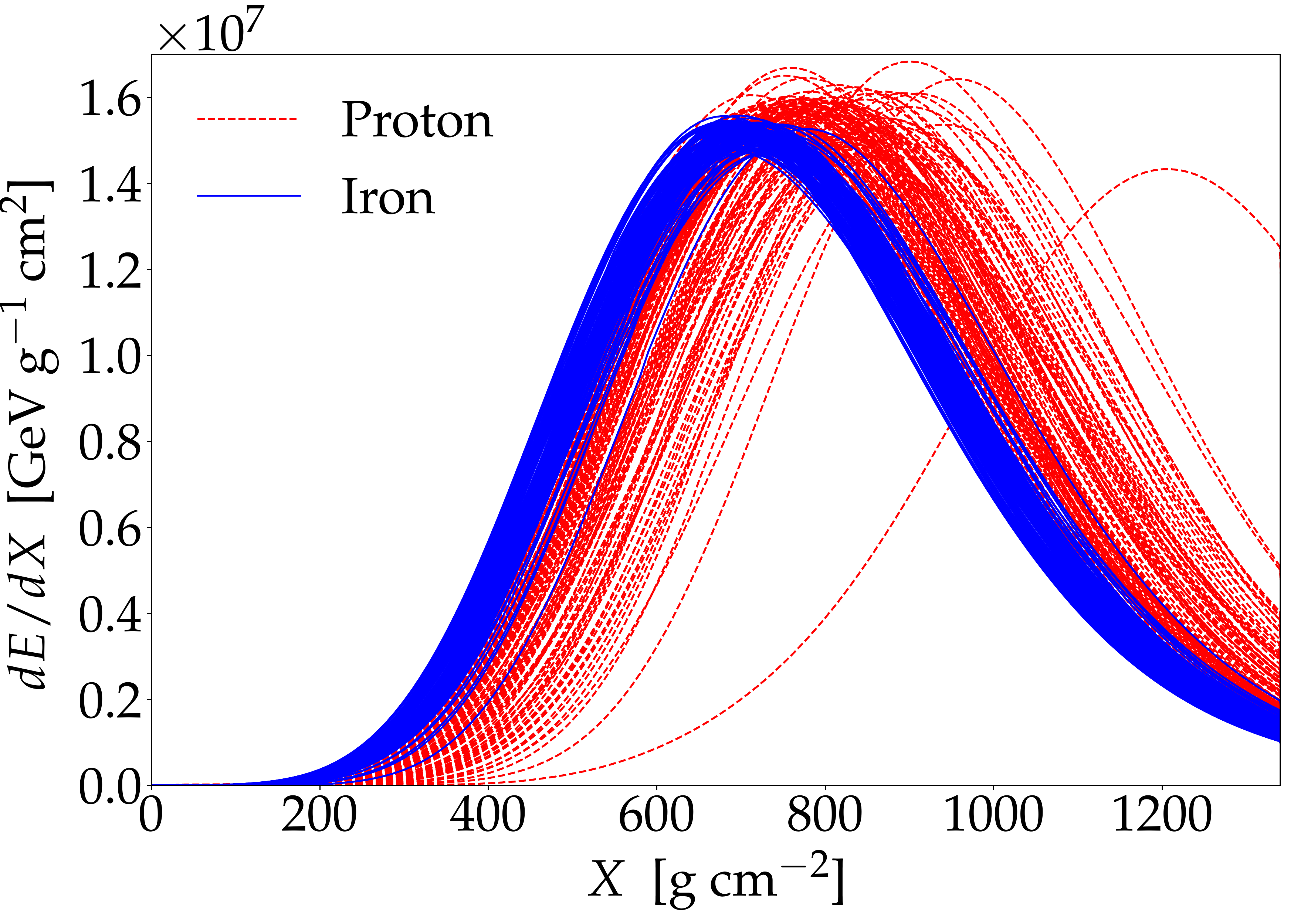}
\includegraphics[width=6.5cm]{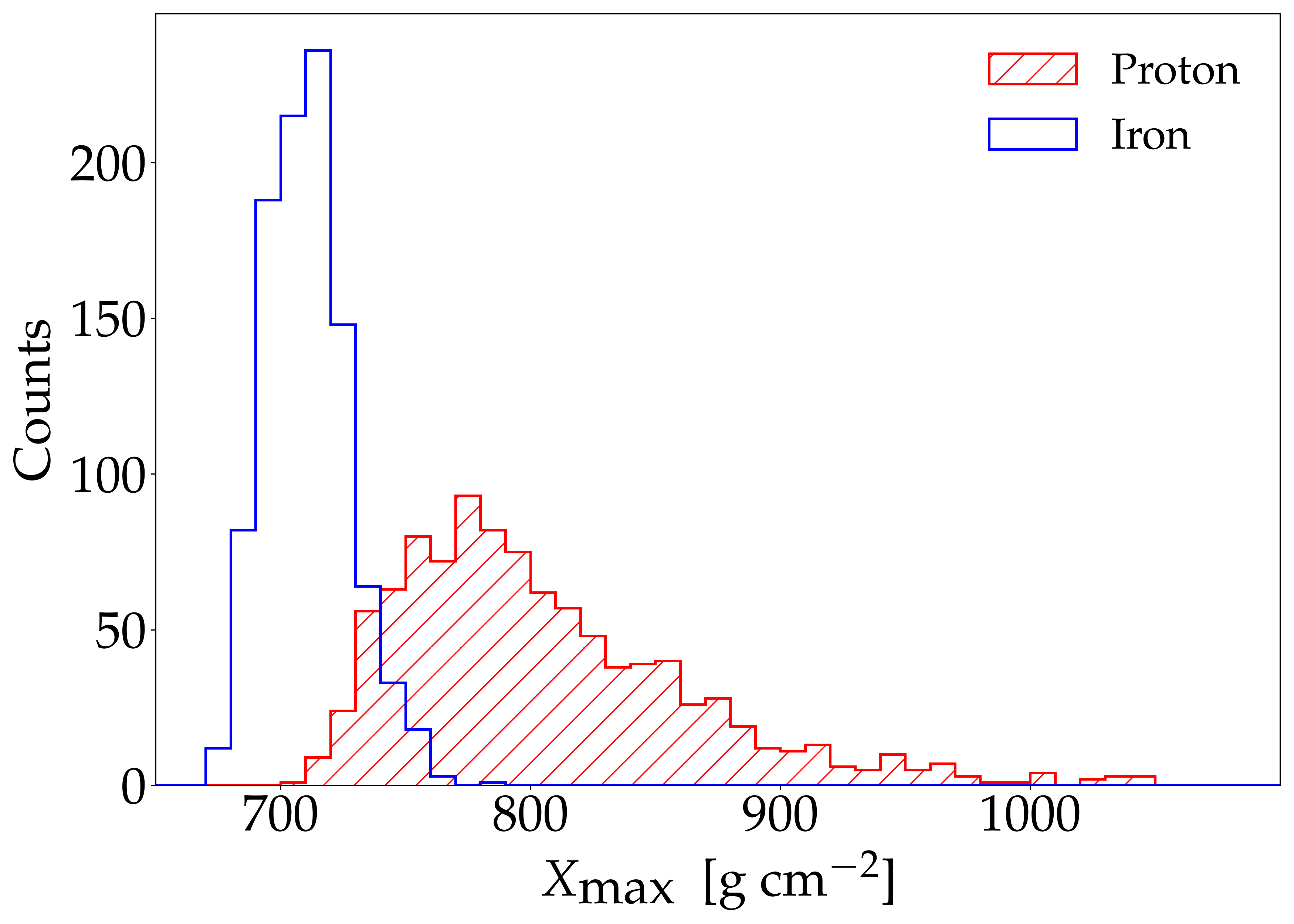}
\caption{\textbf{Left}: Simulated longitudinal profiles for proton- and iron-initiated air showers. \textbf{Righ}: Depth of 
the shower maximum distributions for proton- and iron-initiated air showers. The primary energy is $E=10^{19}$ eV and
the zenith angle is $\theta = 40^\circ$. The simulations are performed with EPOS-LHC. \label{XmaxSims}}
\end{figure}

The left panel of Figure~\ref{XmaxSims} shows the longitudinal profiles of proton and iron showers of $E=10^{19}$ eV,
zenith angle $\theta = 40^\circ$, simulated with CONEX 2r7.5 and EPOS-LHC as the high-energy hadronic interaction model.        
From the figure, it can be seen that effectively the proton showers develop deeper in the atmosphere than the iron showers.
It can also be seen that shower-to-shower fluctuations are larger in the case of proton showers. The right panel of Figure 
\ref{XmaxSims} shows the corresponding distributions of $X_{\textrm{max}}$. Note that both the mean value
and the standard deviation of $X_{\textrm{max}}$ are both sensitive to the primary mass. 

\begin{figure}[ht!]
\centering
\includegraphics[width=6.5cm]{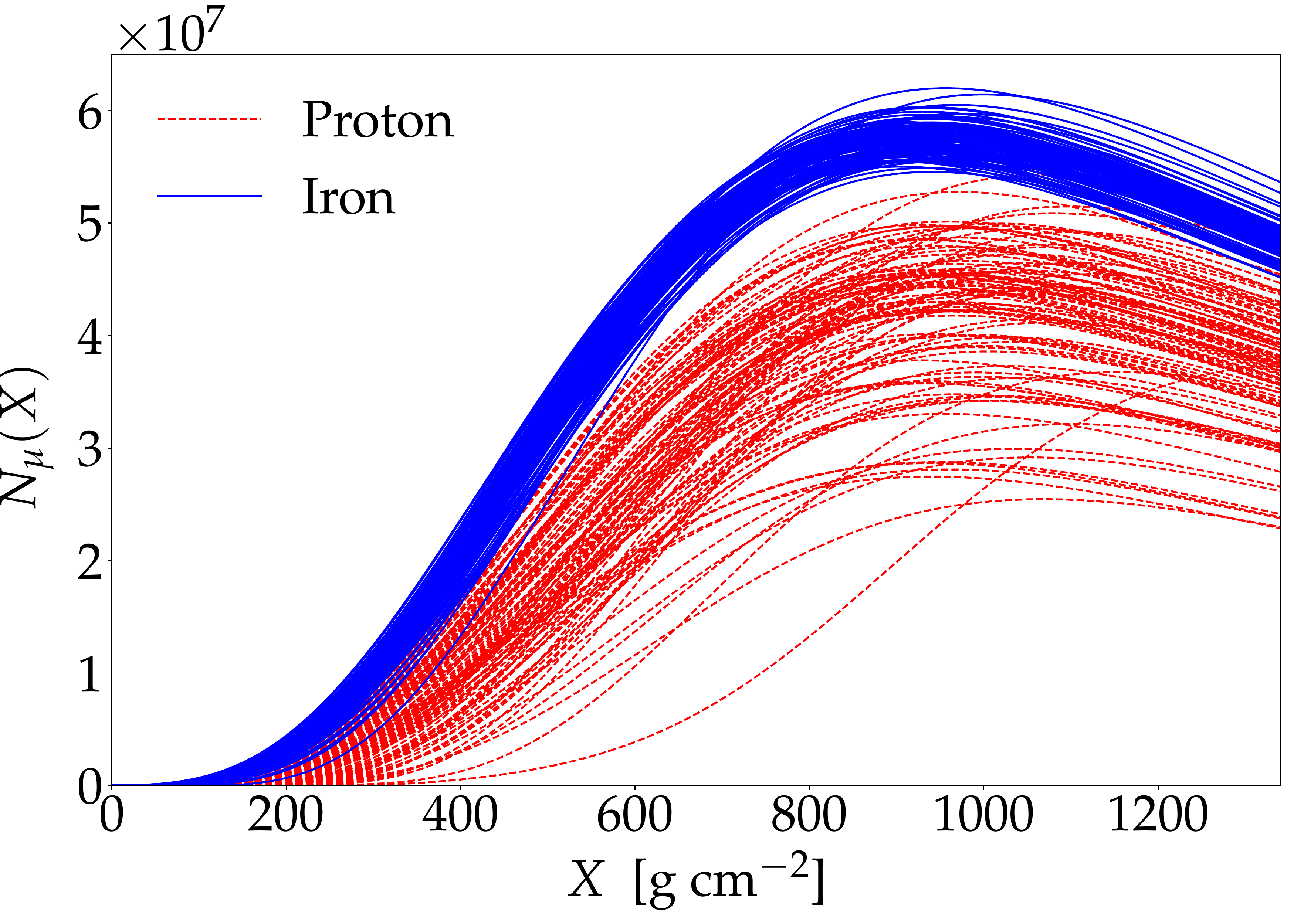}
\includegraphics[width=6.5cm]{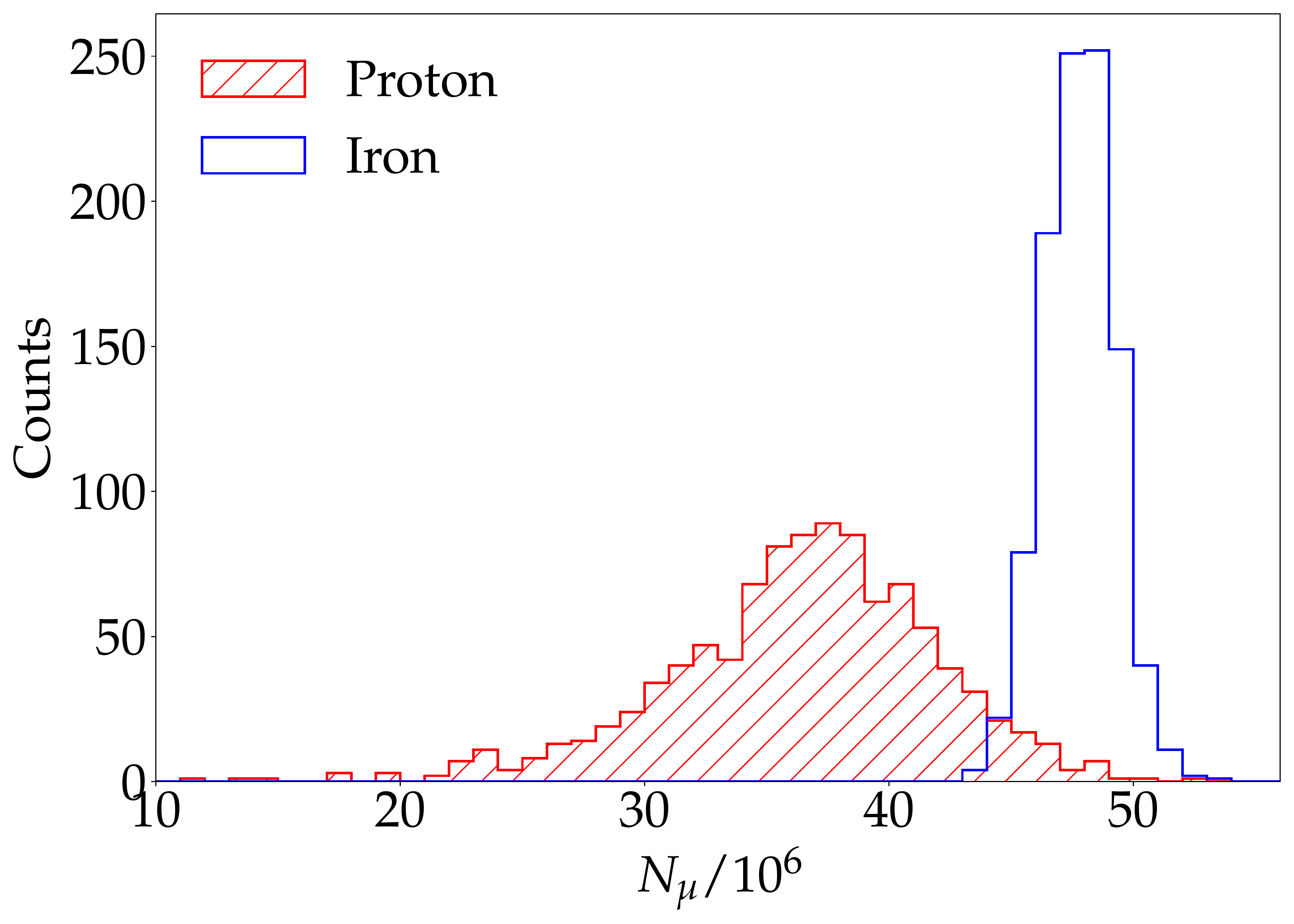}
\caption{\textbf{Left}: Number of muons as a function of the atmospheric depth for proton- and iron-initiated air showers. 
\textbf{Right}: Distributions of the total number of muons at ground for proton- and iron-initiated air showers. The primary 
energy is $E=10^{19}$ eV and the zenith angle is $\theta = 40^\circ$. The observation level is located at sea level 
and the threshold energy of the muons is 1 GeV\@. The simulations are performed with EPOS-LHC. \label{NmuSims}}
\end{figure}

The left panel of Figure~\ref{NmuSims} shows the number of muons, with energy above 1 GeV, as a function of the 
atmospheric depth, for proton- and iron-initiated air showers of \mbox{$E=10^{19}$ eV}, $\theta = 40^\circ$, simulated with 
CONEX 2r7.5 and EPOS-LHC as the high-energy hadronic interaction model. The right panel of the figure shows the 
distributions of the total number of muons at ground. They are obtained evaluating $N_\mu(X)$ at the atmospheric depth 
of the observation level, which is located at sea level. From the figure, it can be seen that the iron showers have a 
larger muon content than proton showers and that the shower-to-shower fluctuations are smaller for iron showers 
as in the case of $X_{\textrm{max}}$. It can also be seen that the total number of muons is also a very good parameter 
to discriminate between proton and iron showers. Note that the number of electrons (hereafter electrons will refer to 
electrons and positrons) of the showers is much larger than the one corresponding to muons. Therefore, the longitudinal 
development of the muons cannot be observed by using optical detectors.    

\begin{figure}[ht!]
\centering
\includegraphics[width=12cm]{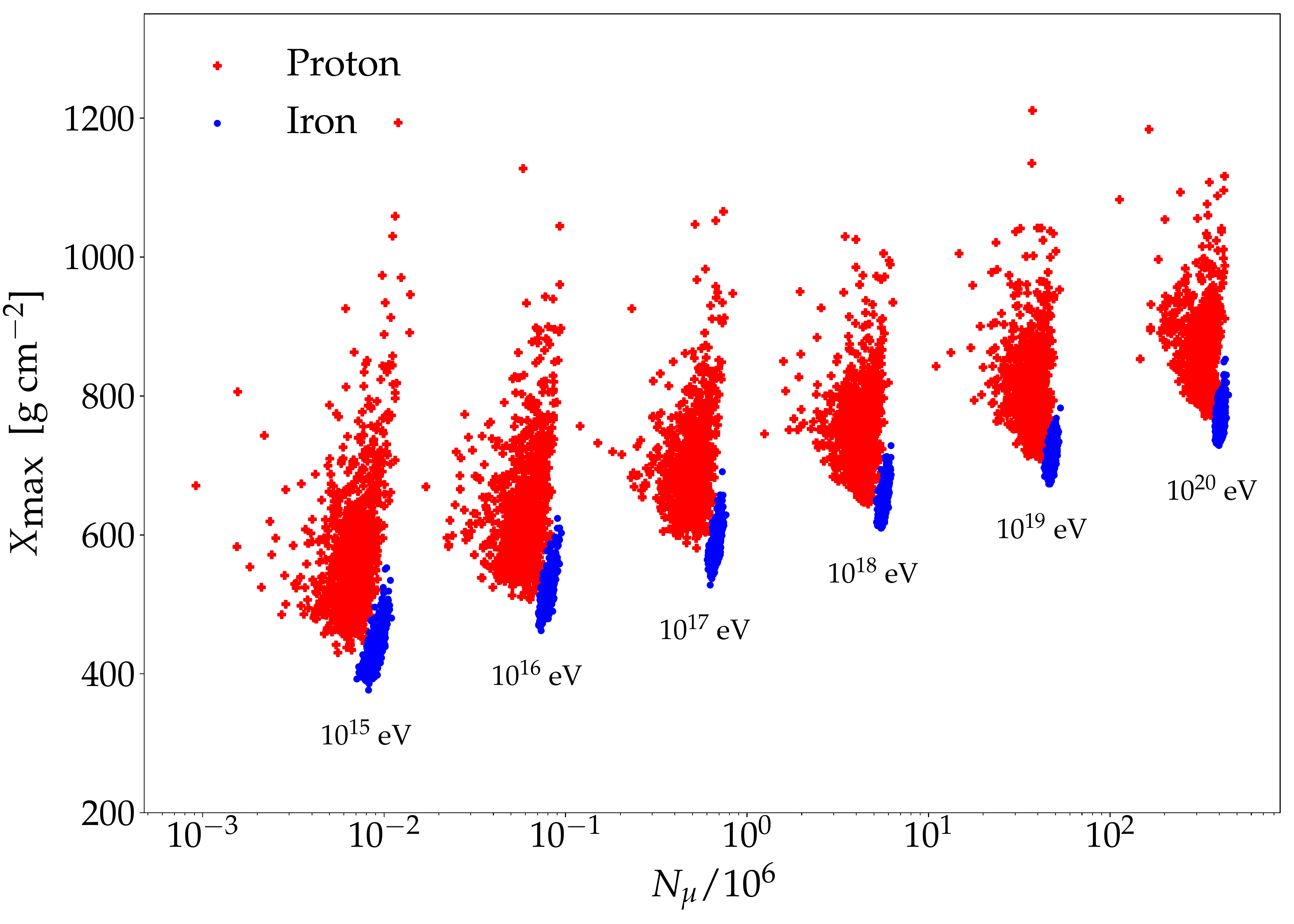}
\caption{Depth of the shower maximum versus the number of muons at ground level for proton- and iron-initiated air 
showers and for different values of primary energy. The zenith angle is $\theta = 40^\circ$. The observation level 
is located at sea level, the threshold energy of the muons is 1 GeV, and the simulations are performed with EPOS-LHC. 
\label{XmaxNmuSims}}
\end{figure}

Figure~\ref{XmaxNmuSims} shows $X_{\textrm{max}}$ as a function of $N_\mu$ for proton and iron showers for different values 
of primary energy. From the figure, it can be seen that these two parameters are very good parameters to separate proton and
iron showers from $10^{15}$  to $10^{20}$ eV, i.e., the whole energy range relevant for cosmic-ray detection through EAS
observations. It can also be seen that the combination of these two parameters increases the discrimination power with 
respect to the case in which any of them is considered individually. This is valid for the whole energy range considered. 

There are many parameters, besides $X_{\textrm{max}}$ and the muon content of the showers, that are sensitive to primary 
mass. Among others are: the slope of the lateral distribution function, the rise time of the signals collected by ground 
detectors, the maximum of the muon production depth distribution, etc. Note that all these parameters depend on 
$X_{\textrm{max}}$ and $N_\mu$. In any case, the discrimination power between heavy and light primaries of the 
$X_{\textrm{max}}$ and $N_\mu$ combination increases very little when more mass-sensitive parameters are added 
\cite{Supanitsky:09b}. 

It is worth mentioning that different cosmic-ray observatories have measured events, different from the vast majority, 
that cannot be explained with current EAS models~\cite{AugerTGF:21,TATGF:20,Beisembaev:21}. These events are characterized 
by large multiplicities of triggered stations and large signals. Auger and Telescope Array have found that these types of 
events are observed during thunderstorms. Even though their origin is still unknown, it is believed that those events may 
originate in bursts of high-energy photons produced during thunderstorms known as downward gamma-ray flashes.

\section{Composition Measurements}
\label{CompoMeas}

In general, there are two types of detectors that allow the detection of EASs: the ones that detect the secondary 
particles of the showers at ground level and the ones that measure the low-energy electromagnetic radiation 
generated due to the shower front propagation through the atmosphere during the shower development. The secondary 
particles at ground are measured with surface detectors and the low-energy radiation produced during the shower 
development is measured with radiation detectors, which include optical and radio detectors.  

\subsection{Composition from Optical and Radio Detectors}

There are two main optical techniques to measure the longitudinal development of the showers, which allow to 
reconstruct the $X_{\textrm{max}}$ parameter directly. These two experimental techniques involve: fluorescence 
telescopes and non-imaging Cherenkov detectors. The radio detectors allow to measure the $X_{\textrm{max}}$
parameter in an indirect way. 

The fluorescence telescopes measure the fluorescence light emitted by the nitrogen molecules of the atmosphere that 
are excited by the charged particles of the EAS. The spectrum of the fluorescence light ranges from 300 to 400 ns. 
The fluorescence light yield is proportional to the energy deposited by the charged particles of the showers in the 
atmosphere. Therefore, measuring the fluorescence light during the shower development, it is possible to reconstruct 
the longitudinal profile of the showers. The fluorescence telescopes are composed by a faceted mirror and 
a camera formed by a large number of small photomultipliers. The telescopes are housed in dedicated climate-controlled 
buildings. The reconstruction of the longitudinal profile requires the determination of the shower axis. It can be 
determined observing the shower development with one telescope (monocular observation), but a much better accuracy 
is achieved when the same shower is observed with two telescopes (stereo observation). A similar accuracy to the one 
corresponding to the stereo observation is obtained when the data of at least one telescope are combined with the data 
taken by surface detectors (hybrid observations). Note that the fluorescence telescopes can take data in clear and 
moonless nights, which reduces the duty cycle to $\sim$$15\%$. Moreover, this experimental technique requires a continuous 
monitoring of the atmosphere.

The fluorescence technique was used in the past in experiments by the Fly's Eye~\cite{flyseye} and its successor, HiRes~\cite{hires}. The Fly's Eye telescope operated 10 years in monocular mode; after that, an additional fluorescence telescope 
was added in order to perform stereo observations. Therefore, in the second phase of the Fly's Eye experiment and during 
the operation of HiRes, it was possible to perform stereo observations. At present, there are two observatories that are
currently taking data and make use of this technique: one is Auger~\cite{Auger} and the other one is Telescope Array
\cite{TA}. Both observatories can perform stereo observations as well as hybrid observations.             

As mentioned before, the energy deposition rate of the showers, $\frac{dE}{dX}$, can be measured by using the 
fluorescence technique. The integral of this profile gives the total energy dissipated electromagnetically.
Therefore, in the fluorescence technique, the atmosphere is used as a calorimeter and the integral of the longitudinal
profile gives the calorimetric energy. The calorimetric energy is $80-90\%$ of the primary energy, and the remaining 
$10-20\%$ corresponds to the so-called ``invisible energy'', which is carried away by neutrinos and high-energy muons 
and has to be estimated and added to the calorimetric energy in order to reconstruct the primary energy (see
Ref.~\cite{AugerIE:19} for details).

\begin{figure}[ht!]
\centering
\includegraphics[width=12cm]{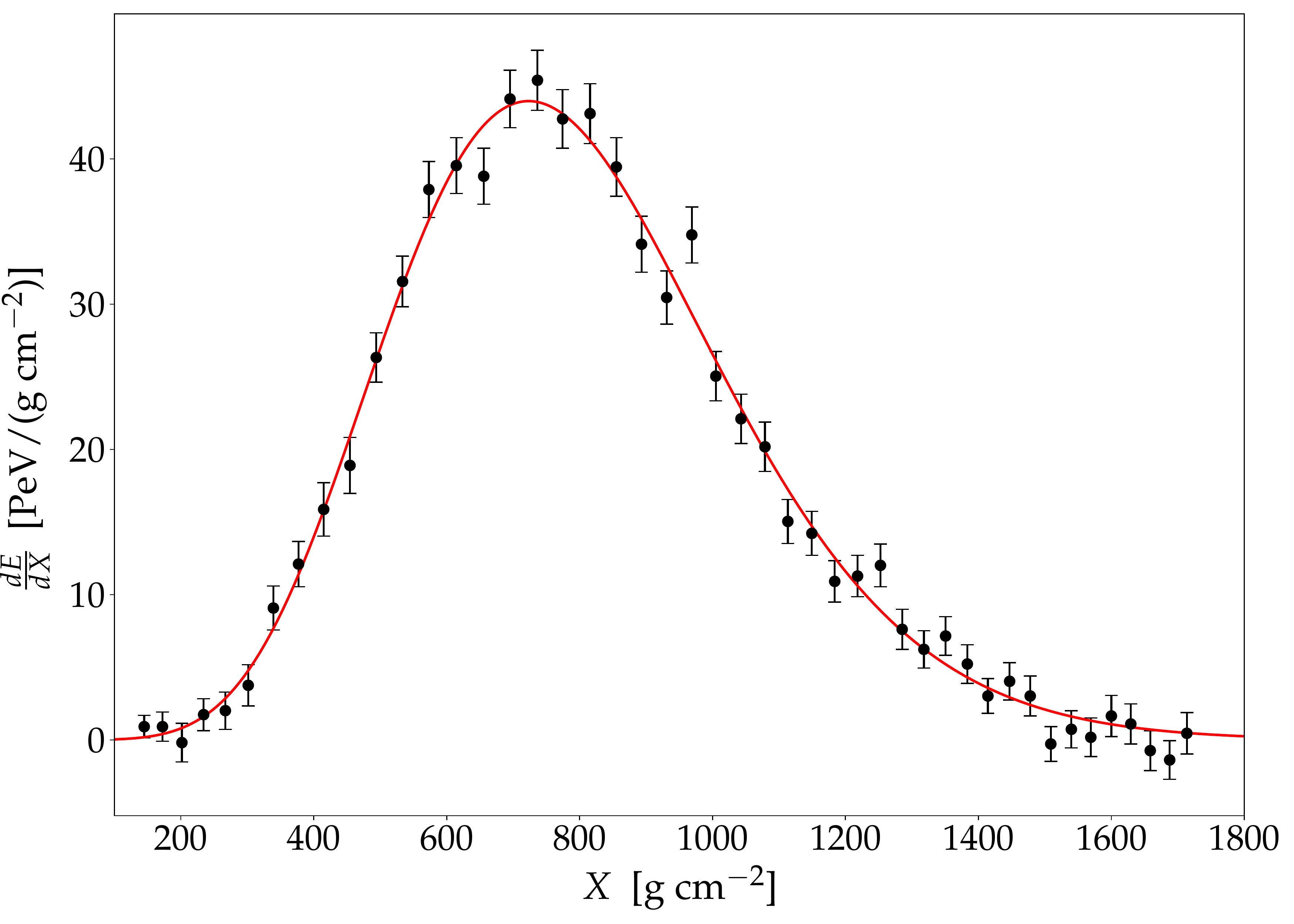}
\caption{Hybrid event detected by the Pierre Auger Observatory. The solid line corresponds to a fit of the data 
points with a Gaisser--Hillas function (see Equation~(\ref{GH})). The primary energy of the event is 
$E = (3.0\pm 0.2)\times 10^{19}$ eV and the maximum is reached at $X_{\textrm{max}}\cong 723$ g cm$^{-2}$. 
Adapted from~\cite{AugerFD:10}. \label{FDEvent}}
\end{figure}

Figure~\ref{FDEvent} shows the longitudinal profile of a hybrid event detected by Auger~\cite{AugerFD:10}. The 
longitudinal profile is fitted with a Gaisser--Hillas function which is given by
\begin{equation}
\label{GH}
f_{\text{GH}}(X) = f_{\textrm{max}} \left( \frac{X-X_0}{X_{\textrm{max}}-X_0} \right)^{\frac{X_{\textrm{max}}-X_0}%
{\lambda}} \exp\left(- \frac{X-X_{\textrm{max}}}{\lambda} \right),
\end{equation}
where $f_{\textrm{max}}$, $X_0$, $\lambda$, and $X_{\textrm{max}}$ are free-fitting parameters. For the event in the
figure, the primary energy is $E_0 = (3.0\pm 0.2)\times 10^{19}$ eV and the depth of the shower maximum is 
$X_{\textrm{max}}\cong 723$ g cm$^{-2}$.

It is worth mentioning that the determination of the primary energy by using the fluorescence technique is 
little influenced by the high-energy hadronic interaction models used to simulate the showers. This is the 
main reason for the hybrid design of Auger and Telescope Array. In contrast with the fluorescence telescopes, 
the duty cycle of the surface detectors is $\sim$$100\%$. Therefore, the calibration in energy of the events 
recorded by the surface detectors only is performed by using hybrid events for which the reconstructed energy is 
taken as the one obtained from the fluorescence telescopes. In this way, the reconstructed energy of the events 
recorded by the surface detectors only are subject to smaller systematic uncertainties compared with the case in 
which the energy calibration is performed by using simulations of the showers.  

The non-imaging Cherenkov detectors measured the Cherenkov radiation that is generated by the charged particles 
of the EAS when they propagate through the atmosphere. Note that the Cherenkov radiation is emitted when a 
charged particle propagates through a medium at a velocity larger than the speed of light in that medium. The 
Cherenkov photons are detected by arrays of photomultipliers (like particle detectors) that look upward. The 
observations are conducted on moonless and clear nights, which reduces the duty cycle to 10--15\%. This technique 
requires a continuous monitoring of the atmosphere. Several cosmic-ray observatories applied the non-imaging 
Cherenkov technique to detect air showers. The three non-imaging Cherenkov detector arrays that are currently 
in operation are the ones installed in Yakutsk~\cite{Ivanov:09}, Tunka~\cite{Budnev:05}, and in the Telescope 
Array site called NICHE~\cite{TANICHE:19}.

\begin{figure}[ht!]
\centering
\includegraphics[width=6.5cm]{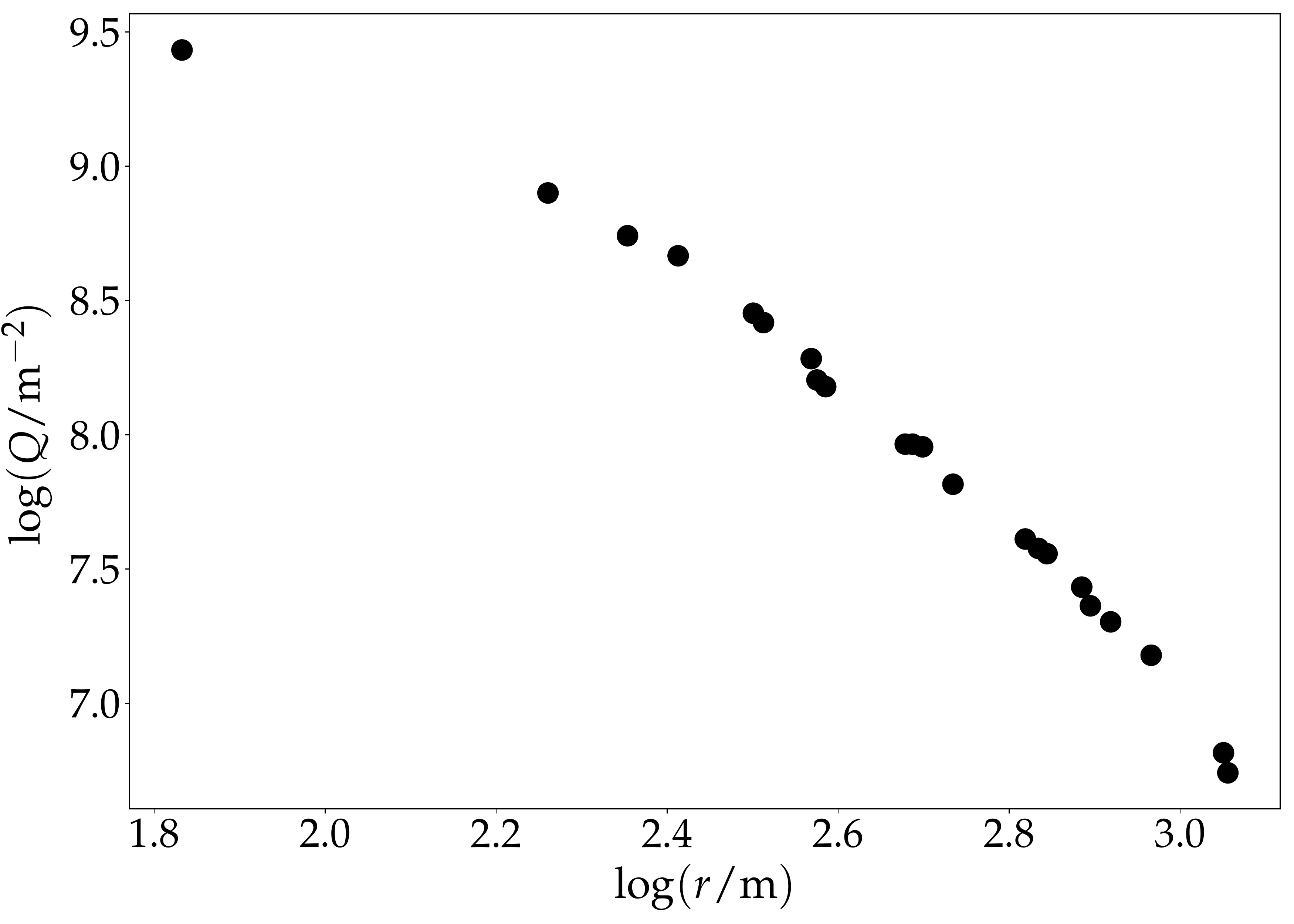}
\includegraphics[width=6.5cm]{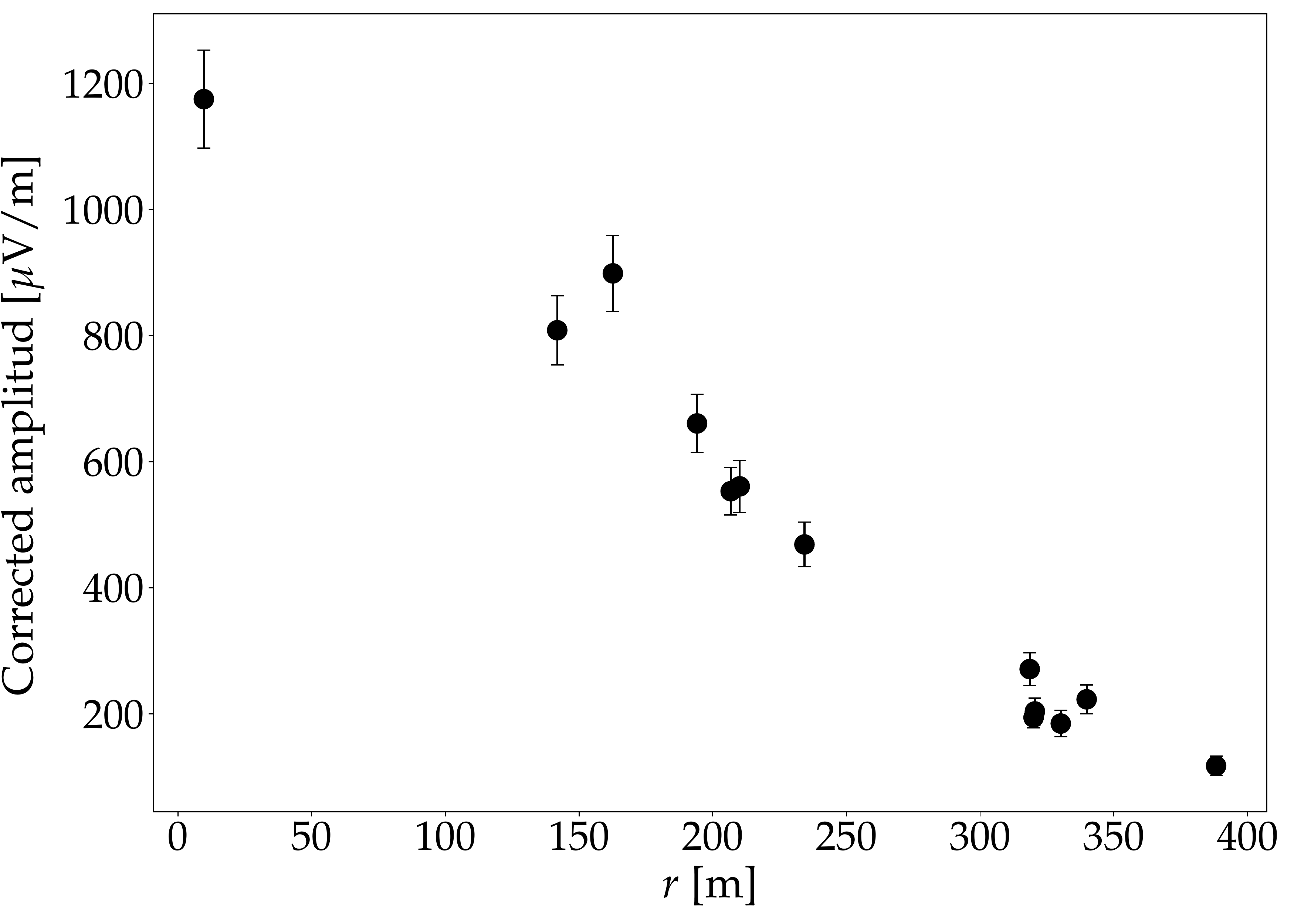}
\caption{\textbf{Left}: Logarithmic photon density as function of the logarithmic distance to the shower axis 
(lateral distribution function) of an event measured by the non-imaging Cherenkov detectors of Yakutsk. The 
primary energy and zenith angle of the event are: $E \cong 1.3 \times 10^{19}$ eV and $\theta \cong 25^\circ$, 
respectively. Adapted from~\cite{Knurenko:19}. \textbf{Right}: Lateral distribution function, after asymmetry 
correction, of an event measured by the Tunka radio antennas in the 35--76 MHz band. Adapted from~\cite{TunkaRex:16}. \label{ChRadio}}
\end{figure}

\textls[-15]{The primary energy and the depth of the shower maximum of each detected shower are obtained from the measured 
lateral distribution function of Cherenkov photons~\cite{Ivanov:09,Dyakonov:86,Budnev:05}.} The left panel of 
Figure~\ref{ChRadio} shows the lateral distribution of Cherenkov photons corresponding to an event measured by 
the Yakutsk {array}~\cite{Knurenko:19}. The primary energy and zenith angle of the event are 
$E \cong 1.3 \times 10^{19}$ eV and $\theta \cong 25^\circ$, respectively. The reconstructed depth of the 
shower maximum is $X_{\textrm{max}}\cong 738$ g cm$^{-2}$. Even though the methods to reconstruct the primary 
energy and $X_{\textrm{max}}$ can be quite sophisticated, it is known that the primary energy is nearly 
proportional to the photons' density at 120 m from the shower axis, and the depth of the shower maximum is 
closely related to the slope of the lateral distribution function~\cite{Hillas:82,Patterson:83}. Note that 
this technique also allows a calorimetric estimation of the primary energy.   

Air showers emit electromagnetic radiation at radio frequencies. This radiation is strongly beamed in the 
forward direction. The two main mechanisms for the radio emission are geomagnetic and Askaryan emissions 
\cite{Huege:16,Schroder:17}. The Cherenkov emission is also present but negligible at radio frequencies. 
The dominant mechanism is the one associated to the geomagnetic field. In this case, the radiation originates 
from the interaction of the secondary electrons and positrons of the showers with the geomagnetic field, 
inducing a time-dependent transverse current. On the other hand, the Askaryan or charge excess emission 
is produced when the air shower particles ionize the atmosphere, and the ionization electrons are added to 
the cascade producing a negative charge excess located in the shower front region. Heavy and positive ions 
remain behind the shower front. In this mechanism, the time-dependent charge excess is responsible for the 
radio emission.

The radio emission from the EAS is emitted in a wide frequency interval. It has been measured from $\sim$$2$  
to $\sim$$500$ MHz. The radio emission is detected through the ground arrays of radio antennas that can measure
this radiation in different frequency bands. The radio signal measured by the antennas is not symmetric
with respect to the shower axis~\cite{Huege:16,Schroder:17}. These asymmetries depend on the relative 
contribution of the geomagnetic and charge excess emissions. There are several methods to reconstruct
$X_{\textrm{max}}$ from the radio signals~\cite{Huege:16,Schroder:17}. In particular, the slope of the 
lateral distribution function is used to reconstruct $X_{\textrm{max}}$. The right panel of Figure 
\ref{ChRadio} shows the lateral distribution function of radio, after asymmetry correction, of an event 
measured by the Tunka experiment~\cite{TunkaRex:16}.         

It is worth mentioning that the radio technique allows a calorimetric determination of the primary energy, 
and also, the radio antennas have a duty cycle close to $100\%$. The experiments that are operating at present 
and that have measured $X_{\textrm{max}}$ by using the radio technique are: Auger~\cite{AugerR:21}, 
Yakutsk~\cite{YakutskR:19}, Tunka~\cite{TunkaR:18}, and LOFAR~\cite{LofarR:21}.
 
\begin{figure}[ht!]
\centering
\includegraphics[width=12cm]{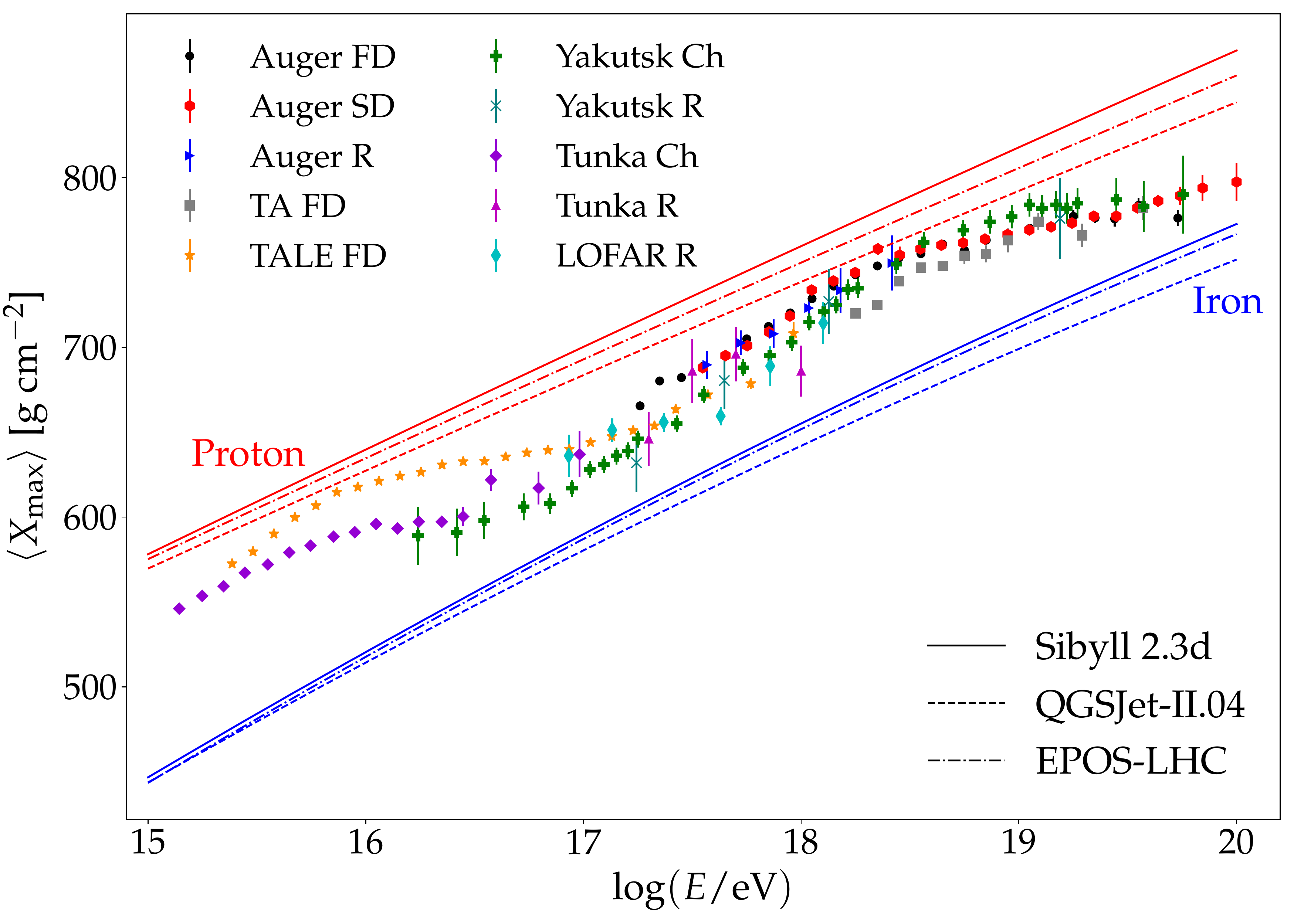}
\caption{Mean value of $X_{\textrm{max}}$ as a function of the logarithmic energy measured by different experiments. 
FD, Ch, and R in the legend refer to fluorescence, Chrerenkov, and radio techniques, respectively. The data of 
Auger are labeled as: Auger FD~\cite{Yushkov:19}, Auger SD (in this case, $X_{\textrm{max}}$ is obtained from 
surface detectors data calibrated with the fluorescence detectors data)~\cite{Peixoto:19}, and Auger R~\cite{AugerR:21}. 
The data of Telescope Array are labeled as: TA FD~\cite{Abbasi:18} and TALE FD~\cite{Abbasi:21}. The data of Yakutsk 
are labeled as: Yakutsk Ch~\cite{Knurenko:19} and Yakutsk R~\cite{YakutskR:19}. The data of Tunka 
are labeled as: Tunka Ch~\cite{Budnev:21} and Tunka R~\cite{TunkaR:18}. The data of LOFAR are labeled as: 
LOFAR R~\cite{LofarR:21}. The curves correspond to model predictions for proton and iron primaries simulated by using 
different high-energy hadronic interaction models.  
\label{MXmaxData}}
\end{figure}

Figure~\ref{MXmaxData} shows the mean value of the depth of the shower maximum as a function of the logarithmic 
primary energy measured by different experiments. The figure also shows the predicted values of 
$\langle X_{\textrm{max}} \rangle$ for proton and iron primaries obtained from shower simulations. The simulated 
air showers used to obtain the model predictions are generated with CONEX 2r7.5 by using the high-energy hadronic 
interaction models: Sibyll 2.3d, QJSJet-II.04, and EPOS-LHC. The primary energy of the simulated air showers ranges 
from $\log(E/\textrm{eV}) = 15$ to  $\log(E/\textrm{eV}) = 20$ in steps of $\Delta \log(E/\textrm{eV}) = 1$ 
and the zenith is $\theta = 40^\circ$. Finally, the curves in the figure are obtained by fitting the mean value of 
$X_{\textrm{max}}$ as a function of the logarithmic energy with a second-order polynomial in $\log(E/\textrm{eV})$.
Note that the $\langle X_{\textrm{max}} \rangle$ measured by the Telescope Array (TA FD in the figure) is shifted by 
5 g cm$^{-2}$ to take into account the detector effects~\cite{Yushkov:18}.   

From Figure~\ref{MXmaxData}, it can be seen that, at low energies, the TALE data present large differences with the 
Tunka and Yakutsk data. These differences are reduced at energies of the order of and even larger than $10^{17}$ eV\@.   
The TALE data are compatible with a light composition at $\sim$$10^{15.4}$ eV that becomes even lighter for increasing 
energies. At energies close to $10^{16}$ eV, they are compatible with a composition that starts to change, becoming heavier 
as the energy increases. The Tunka data, starting at $\sim$$10^{15.1}$ eV, are compatible with a light composition that 
keeps nearly constant up to energies close to $10^{15.8}$ eV\@. From this point, they are compatible with a composition 
that becomes heavier for increasing values of the energy. The Yakutsk data, starting at $\sim$$10^{16.24}$ eV, are 
compatible with the Tunka data in the overlapping energy range. Therefore, even though the experimental data show a transition 
to heavier primaries from $\sim$$10^{16}$ eV, the differences between experiments suggest the existence of important 
systematic uncertainties in the experimental techniques used to obtain the mean value of $X_{\textrm{max}}$ as a function 
of primary energy.

Figure~\ref{MXmaxData} also shows that the Yakutsk data are compatible with a change in the composition 
profile at $\sim$$10^{17}$ eV\@. At larger values of the primary energy, these data are compatible with a variable 
composition that goes from intermediate mass to lighter nuclei. This change is also observed in the TALE and LOFAR 
data but at higher energies. The Auger data, starting at $10^{17.26}$ eV, are also compatible with a variable 
composition that becomes lighter for increasing values of primary energy. Between $\sim$$10^{17.5}$ and 
$\sim$$10^{18.5}$ eV, the data are compatible with a light composition. At $\sim$$10^{18.5}$ eV, these data are compatible
with a composition that becomes heavier for increasing values of primary energy. Even though there are differences 
between the Yakutsk and Auger data, they show a similar trend. Above $\sim$$10^{18.25}$~eV, the Telescope Array data 
seem to be consistent with a constant and light composition, but it has been shown that it is compatible with Auger 
data considering current statistical and systematic uncertainties~\cite{Yushkov:18}.             

\begin{figure}[ht!]
\centering
\includegraphics[width=12cm]{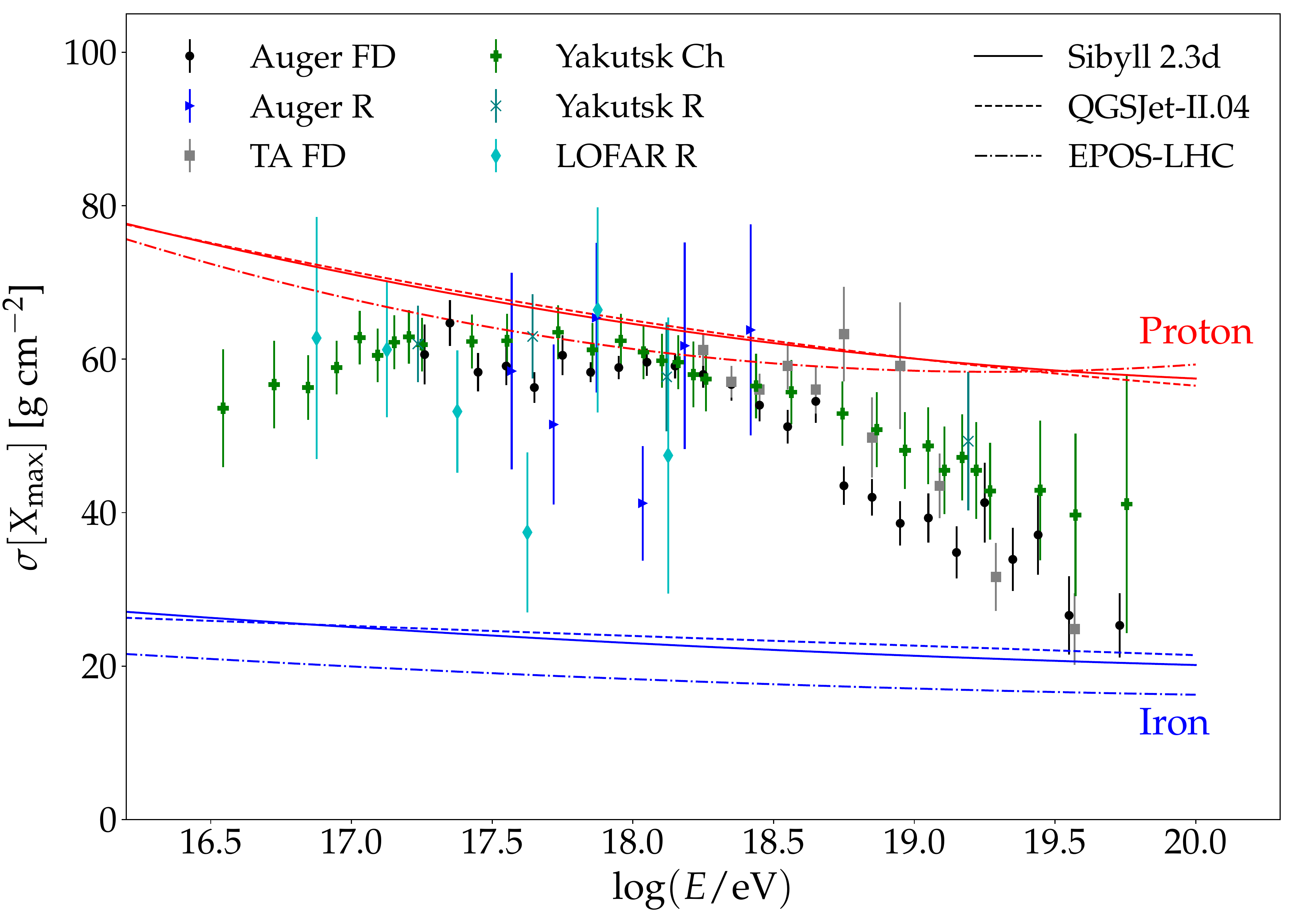}
\caption{Standard deviation of $X_{\textrm{max}}$ as a function of logarithmic energy measured by different experiments.
FD, Ch, and R in the legend refer to fluorescence, Chrerenkov, and radio techniques, respectively. The data of Auger are 
labeled as: Auger FD~\cite{Yushkov:19} and Auger R~\cite{AugerR:21}. The data of Telescope Array are labeled as: TA FD 
\cite{Abbasi:18}. The data of Yakutsk are labeled as: Yakutsk Ch~\cite{Knurenko:19} and Yakutsk R~\cite{YakutskR:19}. 
The data of LOFAR are labeled as: LOFAR R~\cite{LofarR:21}. The curves correspond to model predictions for proton and 
iron primaries simulated by using different high-energy hadronic interaction models.\label{SXmaxData}}
\end{figure}

As mentioned before, the standard deviation of the $X_{\textrm{max}}$ parameter, $\sigma[X_{\textrm{max}}]$, can 
also be used to study the composition of the primary particle. Not all experiments considered have reported the 
$\sigma[X_{\textrm{max}}]$ data. Figure~\ref{SXmaxData} shows $\sigma[X_{\textrm{max}}]$ as a function of the 
logarithmic energy measured by different experiments. The figure also shows the predicted values of 
$\sigma[X_{\textrm{max}}]$ for proton and iron primaries obtained from shower simulations. The same shower library
used to calculate $\langle X_{\textrm{max}} \rangle$ predictions is used to calculate $\sigma[X_{\textrm{max}}]$
predictions. Additionally, in this case, the curves in the figure are obtained by fitting the $\sigma[X_{\textrm{max}}]$ as a 
function of the logarithmic energy with a second-order polynomial in $\log(E/\textrm{eV})$. Note that 
$\sigma[X_{\textrm{max}}]$ measured by the Telescope Array (TA FD in the figure) is obtained by subtracting 15 g cm$^{-2}$ 
in the quadrature to take into account the detector effects~\cite{Batista:19}.    

From Figure~\ref{SXmaxData}, it can be seen that the lowest energy data reported correspond to Yakutsk. They start at 
$10^{16.5}$ eV\@. The Yakutsk data are compatible with a variable composition that seems to be be dominated by 
intermediate nuclei at the lowest energy and becomes lighter for increasing values of primary energy. The data of the 
different experiments are compatible with a light composition from $\sim$$10^{17.5}$ to $\sim$$10^{18.5}$ eV, where a 
transition toward heavier nuclei begins. This picture is compatible with the one inferred considering the mean value 
of $X_{\textrm{max}}$ in the limited energy range corresponding to $\sigma[X_{\textrm{max}}]$ measurements.     

There are several possibilities to study the composition of the cosmic rays from the $X_{\textrm{max}}$ data. From 
the measured $X_{\textrm{max}}$ distributions, it is possible to infer the abundance of different nuclei that are assumed 
to be present in the cosmic-ray energy spectrum at a given energy~\cite{Bellido:18,Abbasi:21}. From the measured values
of $\langle X_{\textrm{max}} \rangle$, it is possible to infer the mean value of the logarithmic mass number 
\cite{AugerXmax:13}, which is defined as $\langle \ln A \rangle = \sum_A f_A(E) \ \ln A$,
%
%
%
%
where $f_A(E)$ is the fraction of nuclei of mass number $A$ at a given primary energy. Moreover, from the measured values 
of $\sigma[X_{\textrm{max}}]$, it is possible to estimate the standard deviation of $\ln A$~\cite{AugerXmax:13}. All 
these analyses have to be conducted by using EAS simulations which, as mentioned before, can introduce important 
systematic uncertainties due to the different predictions obtained when different high-energy hadronic interaction 
models are considered. 

Let us consider the estimation of $\langle \ln A \rangle$ from $\langle X_{\textrm{max}} \rangle$ data. The mean 
value of $X_{\textrm{max}}$ obtained from measurements involves the average over shower-to-shower fluctuation and 
also over the mass number, i.e.,     
\begin{equation}
\label{MXmaxEq}
\langle X_{\textrm{max}} \rangle = \int_0^\infty d X_{\textrm{max}} \sum_A f_A(E) P(X_{\textrm{max}}|A,E)%
\ X_{\textrm{max}},
\end{equation}
where $P(X_{\textrm{max}}|A,E)$ is the distribution function of $X_{\textrm{max}}$ for a given mass number $A$
and primary energy $E$ (see right panel of Figure~\ref{XmaxSims}). Equation (\ref{MXmaxEq}) can also be written 
as,
\begin{equation}
\label{MXmaxEq2}
\langle X_{\textrm{max}} \rangle = \sum_A f_A(E) \langle X_{\textrm{max}}^A \rangle.
\end{equation}

Taking the average over $A$ in Equation~(\ref{XmaxASims}) and using Equation~(\ref{MXmaxEq2}), the following
expression for $\langle \ln A \rangle$ is obtained
\begin{equation}
\label{MLnAEq}
\langle \ln A \rangle = \frac{\langle X_{\textrm{max}} \rangle - \langle X_{\textrm{max}}^\text{p} \rangle}%
{\langle X_{\textrm{max}}^{\text{Fe}} \rangle - \langle X_{\textrm{max}}^\text{p} \rangle} \ln(56),
\end{equation} 
where it is used that the function $F_E$ in Equation~(\ref{XmaxASims}) can be written as 
$F_E=(\langle X_{\textrm{max}}^{\text{Fe}} \rangle - \langle X_{\textrm{max}}^\text{p} \rangle) / \ln(56)$.

\begin{figure}[ht!]
\centering
\includegraphics[width=12cm]{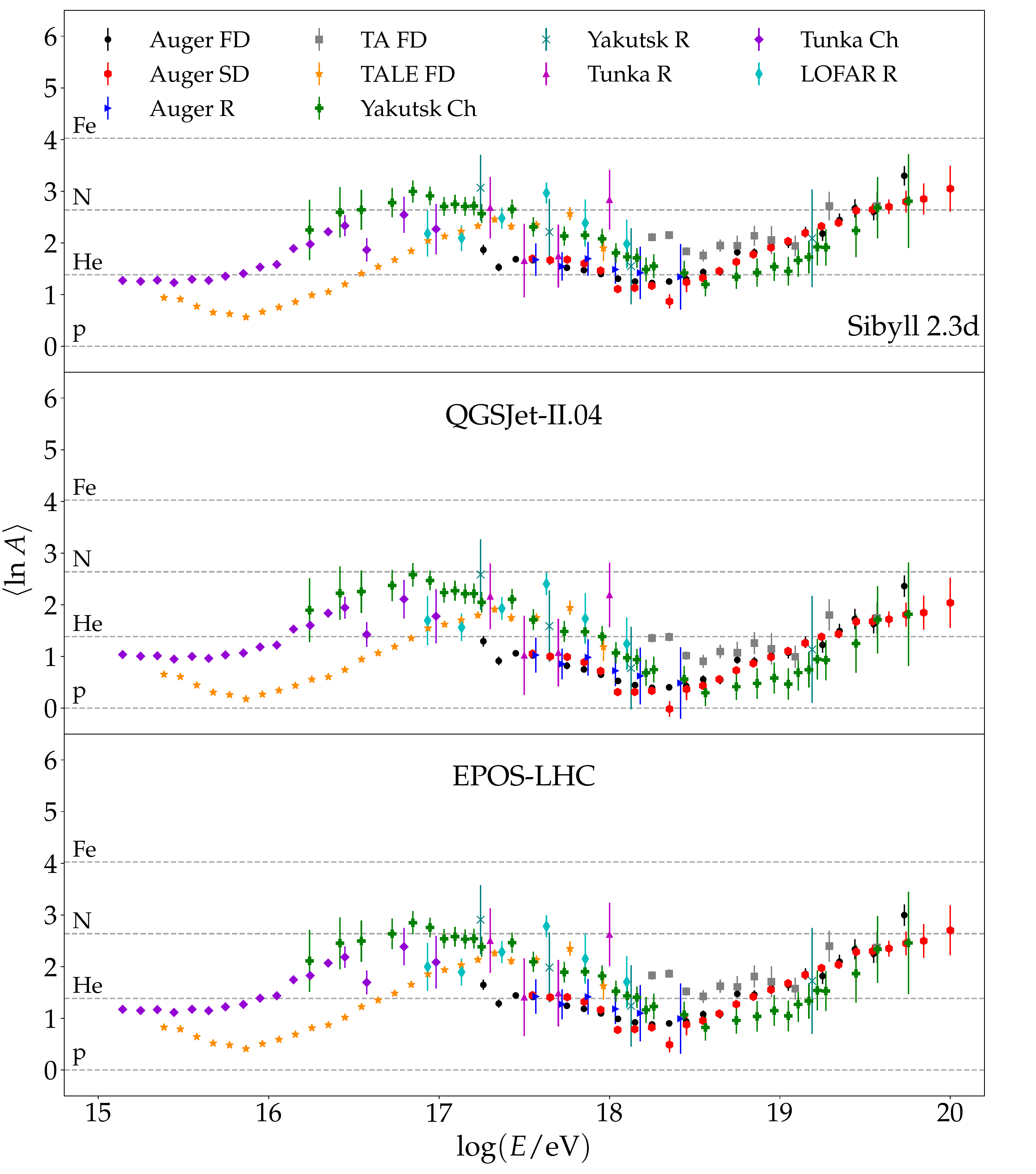}
\caption{Mean value of $\ln A$ as a function of the logarithmic energy obtained from $\langle X_{\textrm{max}} \rangle$ 
measured by different experiments. Three different high-energy hadronic interaction models are used to interpret 
the data. \label{LnA}}
\end{figure}

Figure~\ref{LnA} shows $\langle \ln A \rangle$ as a function of the logarithmic energy obtained by using 
Equation~(\ref{MLnAEq}). The values of $\langle X_{\textrm{max}}^{\text{Fe}} \rangle$ and 
$\langle X_{\textrm{max}}^\text{p} \rangle$ are obtained from the fits of the simulated EAS data described above.
The calculation is performed for the three high-energy hadronic interaction models considered. From the figure, 
the dependence of the inferred $\langle \ln A \rangle$ on the high-energy hadronic interaction 
models used to analyze the data is evident. In fact, the lightest composition is found when QGSJet-II.04 is used to 
analyze the data, and the heaviest composition is found for Sibyll2.3d.

Figure~\ref{LnA} shows more clearly the trend followed by the composition inferred based on Figure~\ref{MXmaxData}. 
For all high-energy hadronic interaction models,  a gradual increase in the composition from 
$\sim$$10^{16}$ eV, reaching a maximum value of $\langle \ln A \rangle$ between $2.6$ and $3$, close to the expectation 
for nitrogen ($A=14$), can be seen. At some point between $\sim$$10^{17}$ and $\sim$$10^{18}$~eV, a transition toward 
light nuclei starts. The minimum is reached at $10^{18.3}$--$10^{18.5}$~eV, where a new transition toward heavy elements
starts. At energies of the order of $\sim$$10^{20}$~eV, the only available $X_{\textrm{max}}$ data are obtained from 
the surface detectors of Auger. This is due to the reduced duty cycle of the fluorescence and non-imaging Cherenkov 
techniques compared with the one corresponding to the surface detectors, which is $\sim$$100\%$. Therefore, at 
$\sim$$10^{20}$ eV, these data suggest that $\langle \ln A \rangle \cong 2-3$ depending on the high-energy hadronic 
interaction model considered.

The change in composition at a break energy of $10^{18.3}$--$10^{18.5}$ eV (the break energy depends on the experimental 
data considered) has been discussed recently in Refs.~\cite{Sokolsky:21,Watson:22}. In these analyses, it is shown that 
there is clear experimental evidence about a change in the elongation rate at a given break energy, as first reported 
by Auger~\cite{Auger:14}. Moreover, in Ref.~\cite{Sokolsky:21}, it is claimed that for energies below the break energy, the 
elongation rate obtained from the data of the northern and southern hemispheres are compatible, but the ones obtained 
considering data above the break energy are different. In contrast, in the subsequent analysis of Ref.~\cite{Watson:22}, 
it is found that the elongation rates measured in the northern and southern hemispheres are compatible below and above 
the energy break.

\subsection{Composition from Surface Detectors}

As mentioned before, the composition can also be inferred from data taken by surface detectors that measure the secondary
particles of the EAS that reach the ground. These composition analyses are based on a single parameter or on a combination 
of several parameters. The main advantage of the composition analyses based on surface detectors data is that the duty cycle 
of these types of detectors is $\sim$$100\%$ and then the number of events is about one order of magnitude larger than the 
one corresponding to the optical detectors.  

In general, it is difficult to compare the results of the analyses based on surface detectors data with the ones obtained 
from the mean value of the $X_{\textrm{max}}$ parameter due to the fact that $\langle \ln A \rangle$ is not always reported 
and it is not directly obtained from the data as in the $X_{\textrm{max}}$ parameter case. In this paper, the results of the 
composition analyses in which $\langle \ln A \rangle$ is reported and the ones corresponding to the measurements of the 
muon density are considered. 

The muon density measured by different experiments cannot be compared directly as it depends on the altitude of the
observatory, the threshold energy of the muons, and the single distance to the shower axis or interval of distances to
the shower axis considered for its calculation. One way to compare measurements of the muon density from different 
experiments is through the $z$-scale~\cite{Dembinski:19,Gesualdi:21b}. However, in this work, a different approach is 
followed. Assuming that the muon density, for a primary of mass number $A$, measured by the different experiments, follows 
Equation~(\ref{NmuANmuPr}), i.e., $\langle \rho_\mu^A \rangle = A^{1-\beta} \langle \rho_\mu^\text{p} \rangle$, and that the mass 
number of the heaviest nuclei is $A_\text{M}$, the $\langle \ln A \rangle$ can be approximated by~\cite{Supanitsky:22}
\begin{equation}
\label{LnASDEq}
\langle \ln A \rangle \cong \left( \frac{\langle \rho_\mu \rangle}{\langle \rho_\mu^\text{p} \rangle} -1 \right) \frac{\ln A_\text{M}}%
{A_\text{M}^{1-\beta}-1} + \frac{\Delta}{2}, 
\end{equation} 
where $A_\text{M}$ is taken as the minimum mass number such that the data points corresponding to a given muon density-related parameter are contained between the predictions for protons and the one corresponding to nuclei of mass number 
$A_\text{M}$ and 
\begin{equation}
\Delta = \left| \frac{A_\text{M}^{1-\beta}-1-(1-\beta)\ln A_\text{M}}{(A_\text{M}^{1-\beta}-1) (1-\beta)}-%
\frac{1}{1-\beta} \ln\left( \frac{A_\text{M}^{1-\beta}-1}{(1-\beta) \ln A_\text{M}}\right) \right|.  
\end{equation}
Here, $\Delta/2$ is the systematic uncertainty introduced by the use of the approximated expression in Equation~(\ref{LnASDEq}).  

\begin{figure}[ht!]
\centering
\includegraphics[width=12cm]{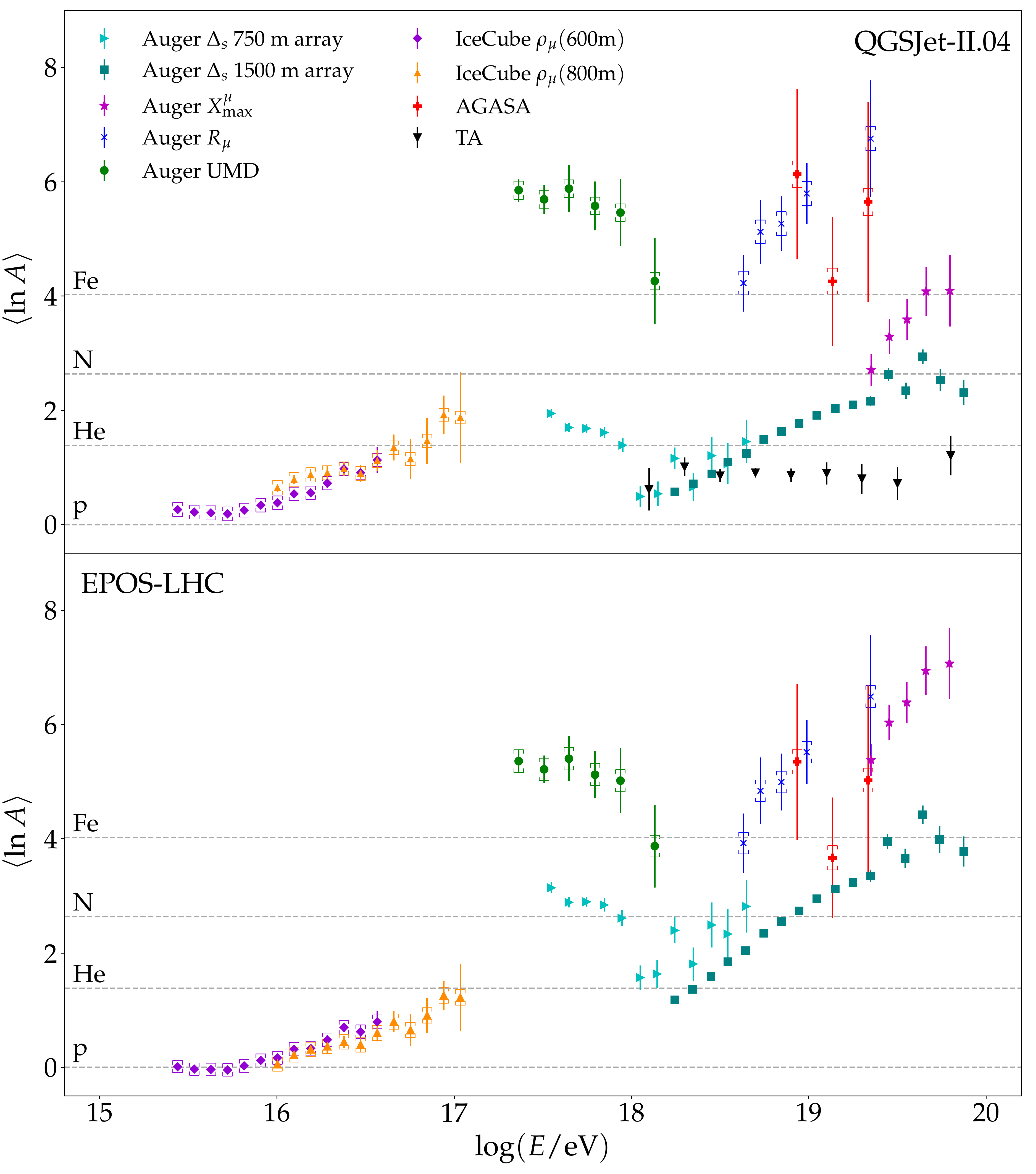}
\caption{Mean value of $\ln A$ as a function of the logarithmic energy obtained from experimental data
taken by the surface detectors of different experiments by using two different high-energy hadronic 
interaction models to interpret the data. The Auger data are labeled as: Auger $\Delta_\text{s}$ 750 m 
array~\cite{AugerDeltaS:17}, Auger $\Delta_\text{s}$ 1500 m array~\cite{AugerDeltaS:17}, Auger $X_{\textrm{max}}^\mu$~\cite{AugerXmumax:14}, Auger $R_\mu$~\cite{AugerRmu:16}, and Auger UMD~\cite{AugerUMD:20}. The IceCube 
data are labeled as: IceCube $\rho_\mu$ (600 m) and IceCube $\rho_\mu$ (800 m)~\cite{IceCube:22}. The AGASA 
data are labeled as: AGASA~\cite{Gesualdi:20,Gesualdi:21}. The Telescope Array data are labeled as: TA~\cite{TA:21}. 
The square brackets correspond to the systematic uncertainties introduced by the use of Equation~(\ref{LnASDEq}) to 
calculate the~$\langle \ln A \rangle$. \label{LnASD}}
\end{figure}

Figure~\ref{LnASD} shows $\langle \ln A \rangle$ as a function of the logarithmic primary energy obtained by using data
from surface detectors. The experiments considered are Auger, IceCube, AGASA, and Telescope Array. In this case, the models
considered are EPOS-LHC and QGSJet-II.04. The measurements considered are: 
\begin{itemize}

\item $\langle \ln A \rangle$ obtained from the $\Delta_\text{s}$ parameter, measured by Auger, which is obtained from 
the risetimes of the signal collected by the water-Cherenkov detectors~\cite{AugerDeltaS:17}. The data taken by the 
750 and 1500 m arrays are considered. 

\item $\langle \ln A \rangle$ obtained from the parameter $X_{\textrm{max}}^\mu$, measured by Auger, which corresponds to
the atmospheric depth of the maximum of the muon production depth distribution~\cite{AugerXmumax:14}. It is obtained from 
the time traces measured by the water-Cherenkov detectors. 

\item The parameter $R_\mu$, measured by Auger, is an estimator of the total number of muons of energy above $0.3$ GeV, 
obtained from the data provided by the water-Cherenkov detectors~\cite{AugerRmu:16}. The events considered correspond 
to inclined showers and are detected in hybrid mode. 

\item The muon density at 450 m from the shower axis with a muon threshold energy around $1$ GeV, measured by the 
Underground Muon Detectors (UMDs) of Auger~\cite{AugerUMD:20}. 

\item The density of GeV muons at 600 and 800 m from the shower axis measured by IceCube with the IceTop 
Array~\cite{IceCube:22}. 

\item The density of muons of energy above $0.5$ GeV evaluated at 1000 m from the shower axis measured by AGASA 
\cite{Gesualdi:20,Gesualdi:21} .    

\item $\langle \ln A \rangle$ obtained through a multiparametric analysis based on the Telescope Array surface 
detectors data~\cite{TA:21}. Note that the $\langle \ln A \rangle$ is reported for the QGSJet-II.04 and 
QGSJet-II.03 (and older versions of the QGSJet-II models) but not for EPOS-LHC.

\end{itemize}       

Note that the energy scales used are the ones corresponding to each experiment, but for AGASA, the energy scale 
from the Spectrum Working Group is used~\cite{SWG:17}. The only systematic uncertainties included in the plot 
are the ones introduced by the use of Equation~(\ref{LnASDEq}) to calculate the $\langle \ln A \rangle$ from 
the measured muon density. 
From Figure~\ref{LnASD}, it can be seen that, in general, the $\langle \ln A \rangle$ obtained from surface detectors
data is incompatible with the one obtained by optical and radio detectors. Moreover, $\langle \ln A \rangle$ obtained 
from the Auger UMD and $R_\mu$ parameter are compatible with a composition dominated by nuclei heavier than iron, 
which is incompatible with any realistic astrophysical scenario. This is so for both EPOS-LHC and QGSJet-II.04. The 
$\langle \ln A \rangle$ obtained from $X_{\textrm{max}}^\mu$ is also compatible with a composition dominated by nuclei 
heavier than iron when EPOS-LHC is considered to analyze the data. When QGSJet-II.04 is considered to interpret the 
$X_{\textrm{max}}^\mu$ data, the $\langle \ln A \rangle$ is compatible with a composition dominated by iron nuclei or 
even lighter for smaller values of the primary energy. However, the values of $\langle \ln A \rangle$ obtained are 
incompatible with the ones obtained by using optical and radio detectors. The AGASA data are also compatible with a flux 
dominated by nuclei heavier than iron. The values of $\langle \ln A \rangle$ obtained by using the $\Delta_\text{s}$
parameter are closer to the ones obtained by using optical and radio detectors, but they are still larger than those. 
The Telescope Array data are compatible with a light composition above $10^{18}$ eV which is also incompatible with 
the optical and radio results and also with other analyses based on surface detectors data. The IceCube data are also 
in tension with the optical results, but in this energy range, there are also discrepancies between data obtained by 
different optical detectors. 

The Auger water-Cherenkov detectors measure muons and also the electromagnetic particles of the showers. Close to the shower
axis, the signal is dominated by the electromagnetic particles, and far from the shower axis, it is dominated by muons. 
Therefore, the $\Delta_\text{s}$ parameter is less affected by the muon component than the other parameters. Because 
the $\langle \ln A \rangle$ obtained from the $\Delta_\text{s}$ parameter is smaller than the one obtained by using 
other parameters completely dominated by the muonic component of the showers, the incompatibility seems to originate 
in the number of muons predicted by the post-LHC high-energy hadronic interaction models. In fact, the much heavier 
composition inferred from muon density measurements suggests that post-LHC high-energy hadronic interaction models 
present a deficit in the muon component~\cite{Albrecht:22}. This muon deficit has been studied in detail by the 
Working Group on Hadronic Interaction and Shower Physics (WHISP). Combining muon data from several experiments, they 
find that the muon measurements are compatible with the post-LHC high-energy hadronic interaction models up to energies 
of a few $10^{16}$~eV\@. Above that energy, the muon deficit increases with the logarithmic energy (see 
Ref.~\cite{WHISP:21} for an updated analysis). The muon deficit can take large values at the highest energies, for 
instance, Auger found that for $10^{18.8}\, \textrm{eV} \le E \le 10^{19.2}\, \textrm{eV}$, it is of the order of 
$\sim$$33\%$ for EPOS-LHC and $\sim$$61\%$ for QGSJet-II.04~\cite{AugerPRL:16}. There is also experimental evidence about 
the increase in the muon deficit with the zenith angle of the showers~\cite{AugerPRL:16,KASCADEG:18}. It is worth 
mentioning that the composition inferred from the muon measurements of the Yakutsk EAS array is compatible with the one 
inferred from the $X_{\textrm{max}}$ measurements~\cite{Yakutsk:19}. This is so in the whole energy range of the Yakutsk 
EAS array, which includes the highest energies where the muon deficit found by other experiments is more important. 
Therefore, further studies of the biases and systematic uncertainties of each experiment are required in order to 
understand and solve the existing discrepancies.

It is worth mentioning that, despite the tension between the experimental data and the post-LHC high-energy hadronic 
interaction models, the trends in the change of the mass composition obtained from optical detectors (see Figure~\ref{LnA})
are consistent with the ones obtained from surface detectors (see Figure~\ref{LnASD}).          

\begin{figure}[ht!]
\centering
\includegraphics[width=6.5cm]{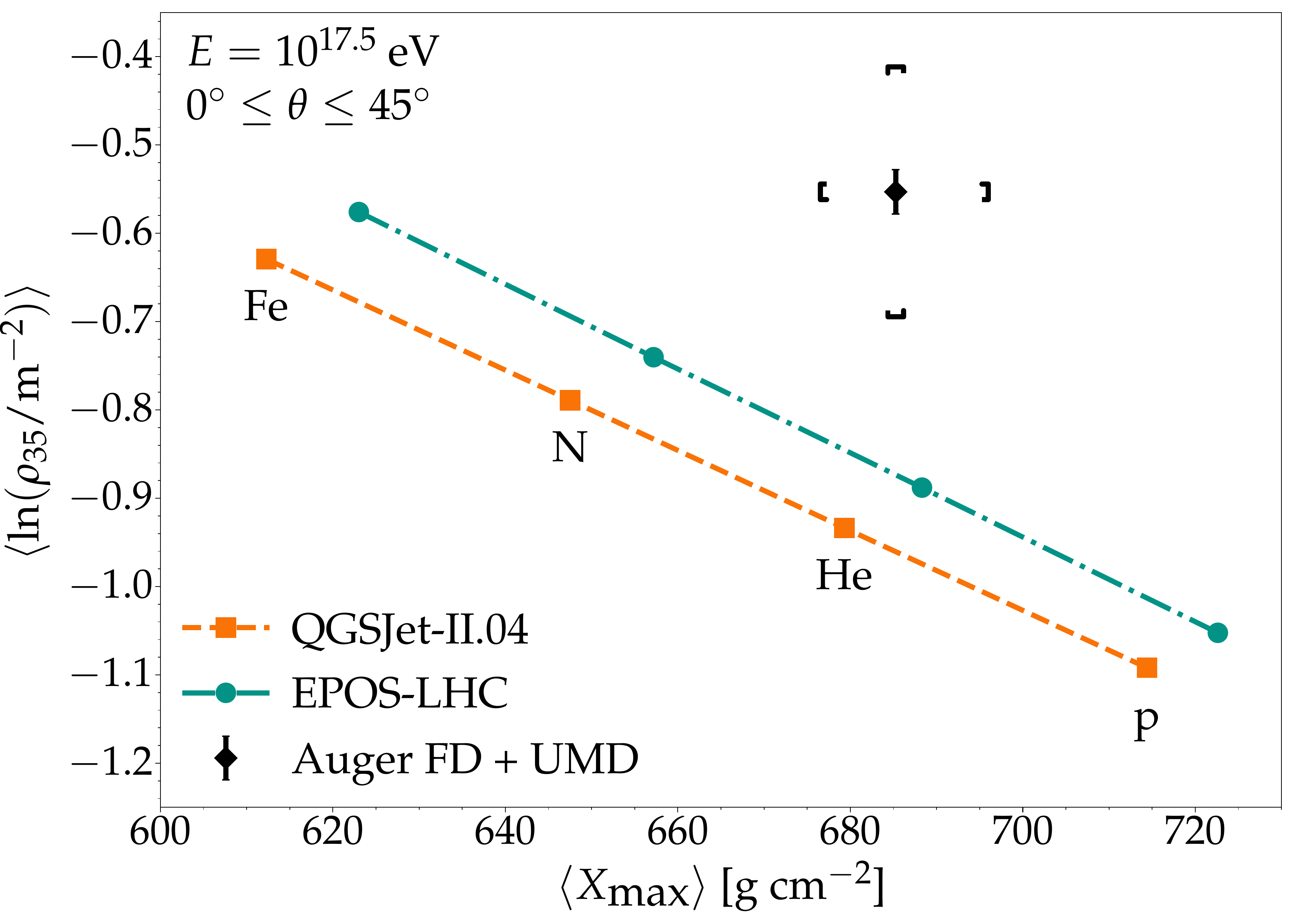}
\includegraphics[width=6.5cm]{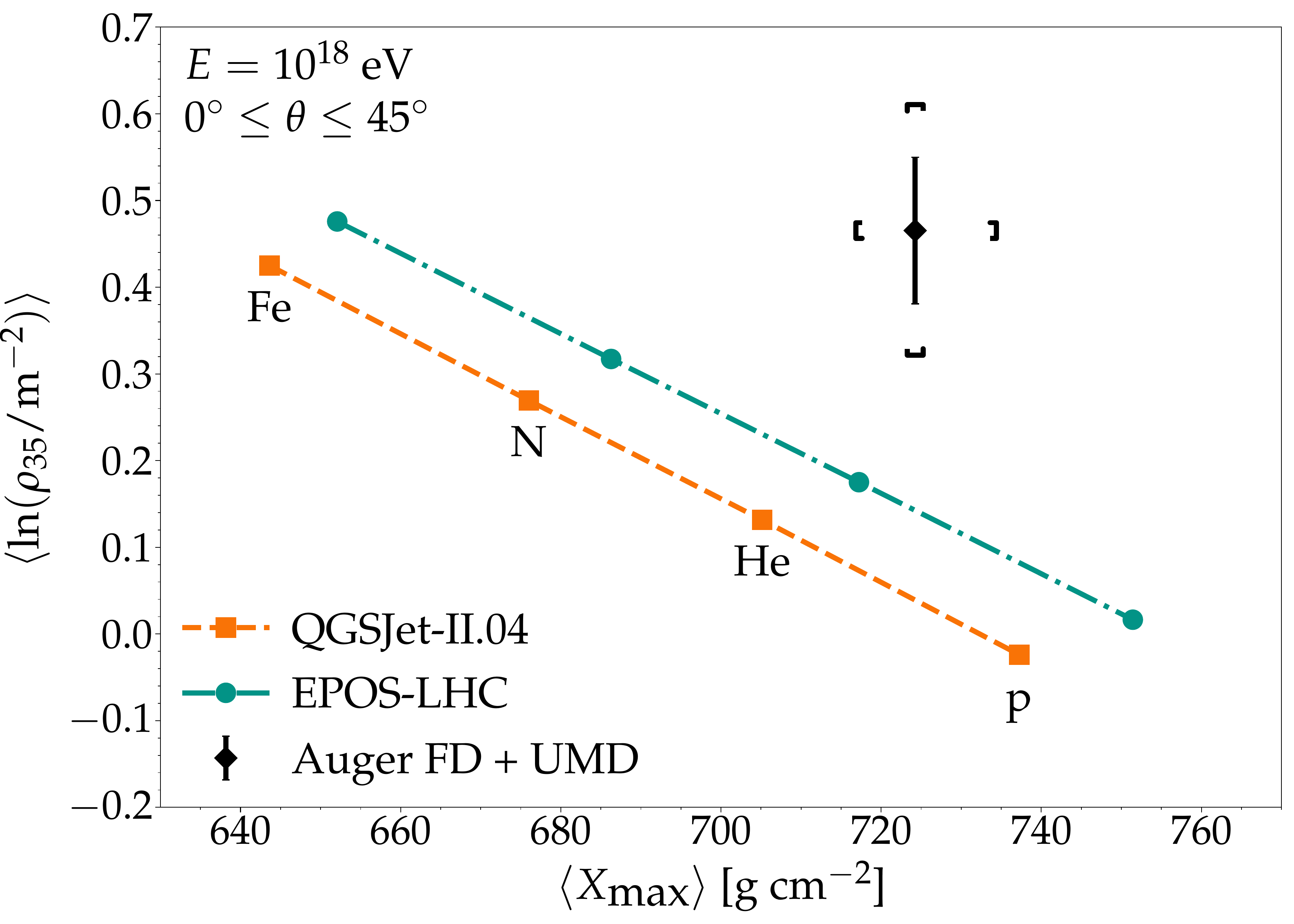}
\includegraphics[width=6.5cm]{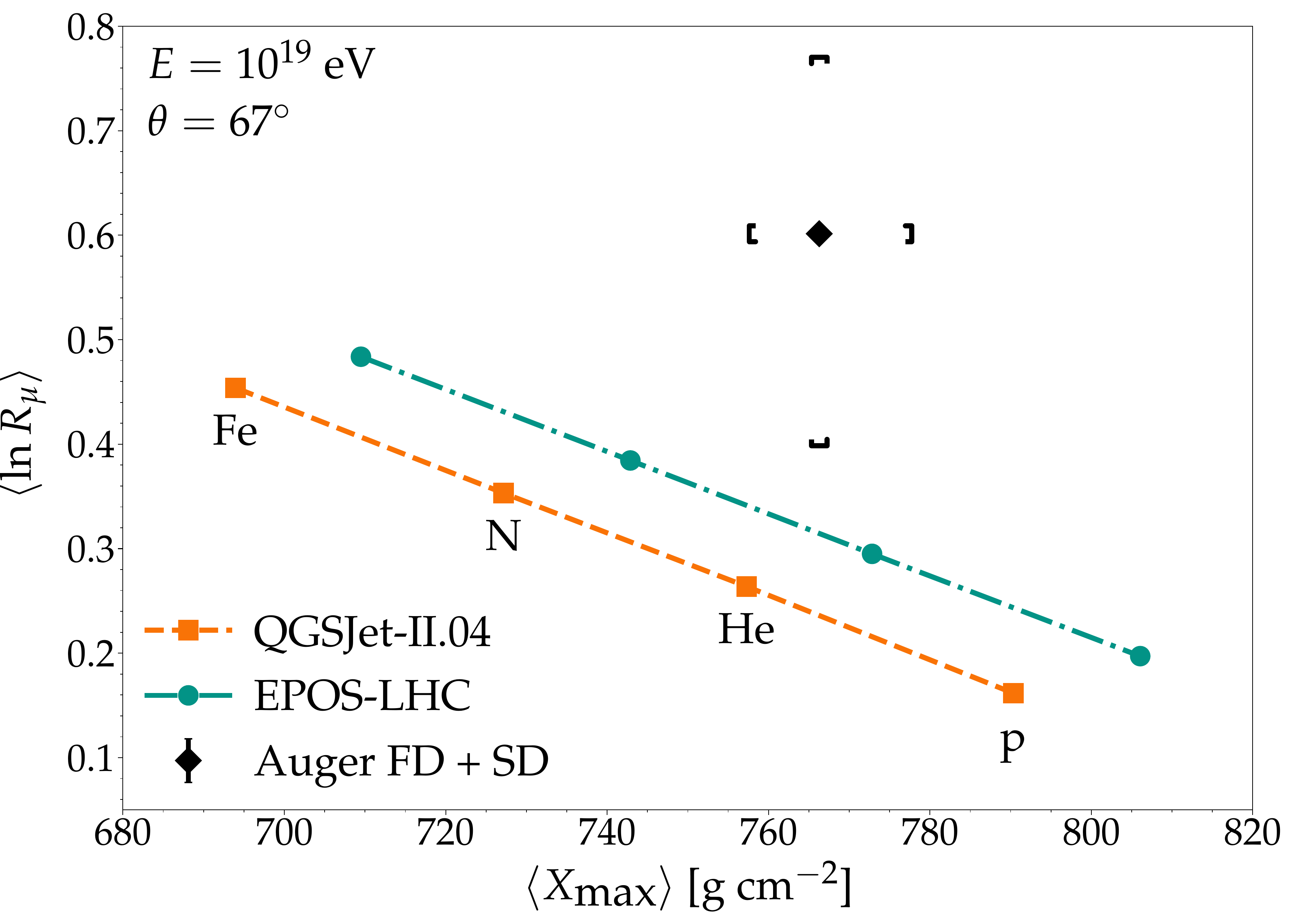}
\includegraphics[width=6.5cm]{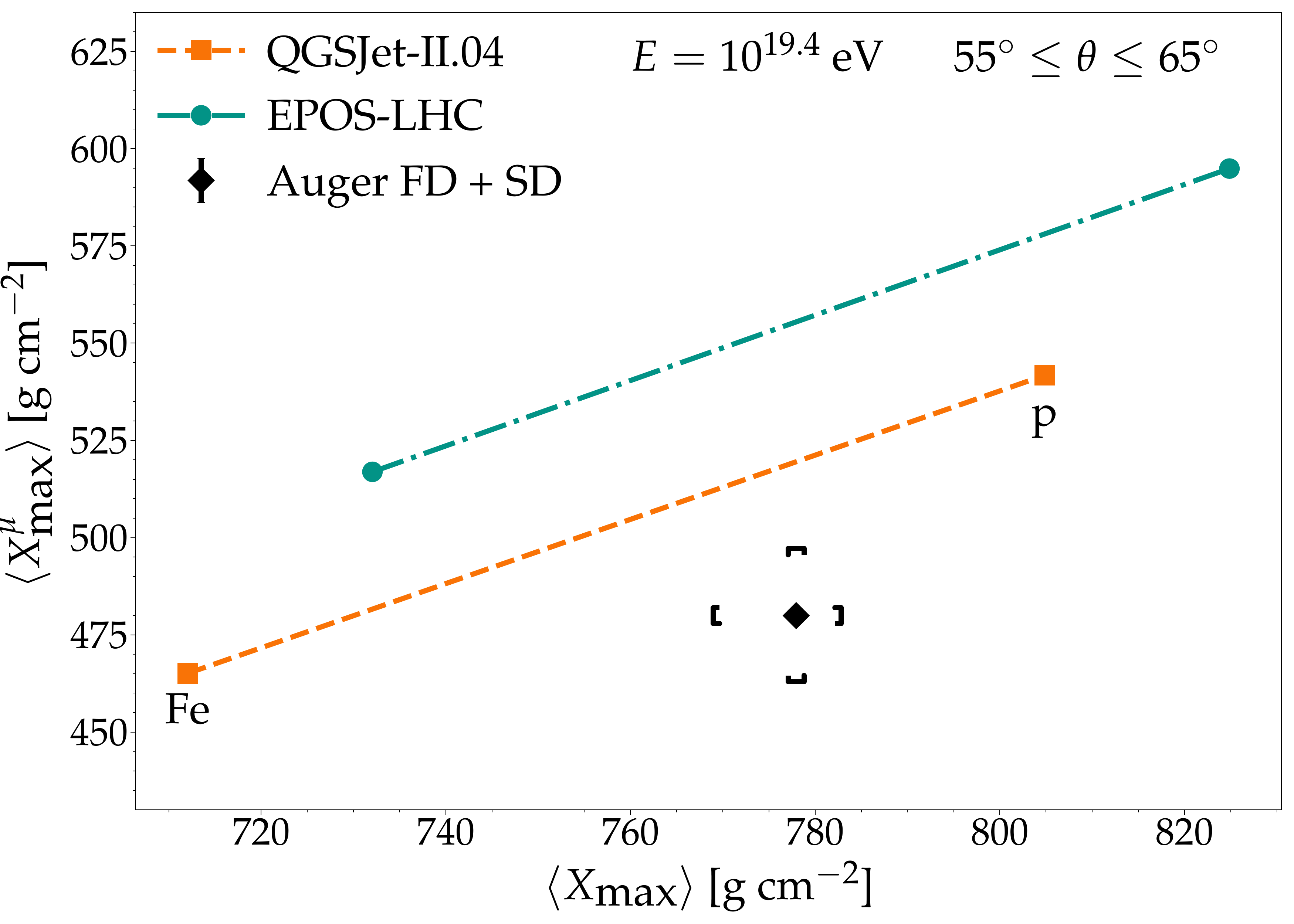}
\caption{Top panels: Mean value of the logarithmic muon density at 450 m from the shower axis for a reference 
zenith angle of $35^\circ$ as a function of the mean value of $X_{\textrm{max}}$ measured by the Auger UMD. 
Adapted from~\cite{AugerUMD:20}. Bottom left panel: Mean value of the natural logarithm of $R_\mu$ as a function 
of the mean value of $X_{\textrm{max}}$ measured by the Auger water-Cherenkov detectors. Adapted from~\cite{AugerRmu:16}.
Bottom right panel: Mean value of $X_{\textrm{max}}^\mu$ parameter as a function of the mean value of $X_{\textrm{max}}$ 
measured by the Auger water-Cherenkov detectors. Adapted from~\cite{Cazon:19}. The curves correspond to the predictions
obtained for EPOS-LHC and QGSJet-II.04. \label{XmaxNmu}}
\end{figure}

In order to illustrate the tension between post-LHC high-energy hadronic interaction models and the experimental data, 
at the level of the composition-sensitive observables, Figure~\ref{XmaxNmu} shows the mean values of different parameters 
measured by using data taken by the UMDs (top panels) and the water-Cherenkov detectors (lower panels) of Auger as a function 
of the mean value of $X_{\textrm{max}}$ for different values of the primary energy and zenith angle. The curves correspond 
to the predictions obtained by using EPOS-LHC and QGSJet-II.04. From the figure, it can be seen that the experimental data 
are incompatible with the predictions of the high-energy hadronic interaction models considered. Note that the experimental
points fall outside the curves even when the systematic uncertainties are considered. Moreover, without considering the analysis
based on the $X_{\textrm{max}}^\mu$ parameter, which is not directly related with the muon content of the showers, it can 
be seen that the tension between high-energy hadronic interaction models considered and data can be relaxed by increasing
the number of muons in the simulated showers.    
In the second knee region, KASCADE-Grande could separate the flux in light and heavy components measuring the 
shower particles with shielded and unshielded surface detectors~\cite{KG:11,KG:13}. Note that with these two types 
of surface detectors, it is possible to separate the muonic and the electromagnetic components. In this case, the 
primary energy is reconstructed by using simulated showers. The heavy component presents a knee-like feature at 
$\sim$$10^{16.7}$--$10^{16.8}$ eV when the data are analyzed with EPOS-LHC, QGSJet-II.04, and Sibyll 2.3d~\cite{KG:21}.
The light component presents a flattening at $\sim$$10^{17}$~eV\@. At this energy, the flux is dominated by the heavy 
component. If the cosmic ray flux in the knee region is dominated by protons, the iron knee should take place at
$E_{\text{Fe}}^{\text{knee}} = 26 \times E_\text{p}^{\text{knee}}$; therefore, if 
$E_\text{p}^{\text{knee}} = 3 \times 10^{15}\, \textrm{eV} \cong 10^{15.48}$ eV, then 
$E_{\text{Fe}}^{\text{knee}} \cong 10^{16.89}$ eV, very close to the knee observed by KASCADE-Grande in the heavy 
component. In this interpretation, the knee in the heavy component is a signature of the end of the galactic cosmic 
rays. The ankle-like feature in the light component can be interpreted as the beginning of the transition between 
galactic and extragalactic cosmic rays~\cite{KG:13}. In this case, this feature is formed by the superposition of a 
galactic flux with the extragalactic one which starts to be dominant above $\sim$$10^{17}$ eV\@.

The increase in the $\langle \ln A \rangle$ measured by TALE above $10^{17}$ eV (see Figure~\ref{LnA}) is in tension 
with this interpretation. Additionally, the iron fraction measured by  TALE~\cite{Abbasi:21} does not show any evidence about the 
knee in the heavy component observed by KASCADE-Grande. Moreover, the $\langle \ln A \rangle$ measured by  IceTop in 
combination with IceCube also increases above $10^{17}$ eV~\cite{IceCubeJ:19}. Moreover, in this case, the iron fraction does 
not show any change below $10^{17}$ eV\@.    

The discrepancies found between data taken by surface detectors in the second knee region can be due to the muon 
deficit of the simulated showers. As mentioned before, the tension among data taken by optical detectors in this 
region of the spectrum suggest the existence of large systematic uncertainties. More studies are required to 
understand the differences found.

\subsection{Combined Analyses}

The results of a study performed combining data taken by the fluorescence telescopes and the water-Cherenkov detectors 
of Auger were recently published in Ref.~\cite{Vicha:21}. The primary energy of the events considered in the analysis 
ranges from $10^{18.5}$ to $10^{19}$ eV, and the parameters considered are $X_{\textrm{max}}$ and the signal deposited 
in the water-Cherenkov detectors at 1000 m from the shower axis, $S(1000)$. Assuming that the composition of the cosmic 
rays is independent of the zenith angle, it was found that to alleviate the tension between the Auger data and EAS 
simulations, in addition to the increase in the number of muons of the simulated showers, a shift in the simulated 
$X_{\textrm{max}}$ parameter is also required. If this result is confirmed in future studies, it would imply that the 
muon deficit alone cannot explain the discrepancies between current high-energy hadronic interaction models and 
experimental data.  
 
A scaled version of the same parameters has been considered by Auger to study the purity of the cosmic rays in the 
energy range from $10^{18.5}$  to $10^{19}$ eV\@. In this case, the correlation between the scaled version of the
parameters is used to study the composition~\cite{Yushkov:19,AugerCorr:16}. For energies between $10^{18.5}$ and 
$10^{18.7}$ eV, a pure composition is excluded with a significance larger than $6.4\, \sigma$. Moreover, the data can 
be explained only by a mixture of primary nuclei of a mass number heavier than helium, i.e., pure composition, and 
proton--helium mixtures are disfavored by the data. An important feature of this analysis is that it is nearly 
independent on the high-energy hadronic interaction models used to analyze the experimental data. Note that this 
result is consistent with the composition inferred from the $X_{\textrm{max}}$ parameter alone, measured by Auger.  

There are two main interpretations about the formation of the ankle and the origin of the light component that dominates
the flux between $10^{18}$ and  $10^{18.5}$ eV\@. In the first scenario~\cite{Aloisio:14,Mollerach:20}, the light component 
below the ankle originates in a different population of sources than the one that dominates the flux above the ankle, 
which includes heavier nuclei according to the Auger data. In the second scenario, the light component originates from 
the photodisintegration of high-energy and heavier nuclei in a photon field present in the sources or its environment
\cite{Unger:15,Globus:15a,Globus:15b,Kachelriess:17,Fang:18,Supanitsky:18}. In these two scenarios, the transition between 
the galactic and extragalactic components takes place in the second knee region.

\section{High-Energy Photon and Neutrino Searches}
\label{PhNu}

Ultra-high-energy photons and neutrinos carry very important information about the cosmic-ray accelerators and the 
propagation of the cosmic rays in the intergalactic medium. In particular, its arrival direction points back to the 
source, as they are neutral particles and then they are not deflected by the galactic and extragalactic magnetic
fields. Moreover, knowing the photon flux level is very important for an accurate energy calibration of surface detectors 
with fluorescence telescopes (see, for instance, Ref.~\cite{AugerPRL:08}). 

As mentioned before, a flux of high-energy photons and neutrinos is expected due to the interaction of the 
extragalactic cosmic rays with the radiation field present in the intergalactic medium. Nuclei of ultra-high energies 
can interact with the low-energy photons of the extragalactic background light, the cosmic microwave background, and 
the radio background. Neutrinos are produced in pion decays, generated through photo-pion production and nuclear decay. 
The main decay channels of charged pions are: 
$\pi^+ \rightarrow \mu^+ + \nu_\mu \rightarrow e^+ +\nu_e + \bar{\nu}_\mu + \nu_\mu$ and 
$\pi^- \rightarrow \mu^- + \bar{\nu}_\mu \rightarrow e^- + \bar{\nu}_e + \nu_\mu + \bar{\nu}_\mu$. Note that the 
muon decay also contributes to the neutrino generation. The expected flavor ratio at Earth is $1:1:1$ due to neutrino 
oscillations. Photons are produced in neutral pion decays ($\pi^0 \rightarrow \gamma + \gamma$), generated in 
photo-pion production, and also due to the inverse Compton of high-energy electrons with low-energy photons of the 
background ($e^\pm + \gamma_b \rightarrow e^\pm + \gamma$). Note that high-energy electrons are generated by the pair 
production of high-energy nuclei with the low-energy photons of the background. They are also created in the beta decay 
of nuclei and the decay of muons from the decay of charged pions. The high-energy photons and electrons form an 
electromagnetic cascade in the intergalactic medium.     

\begin{figure}[ht!]
\centering
\includegraphics[width=12cm]{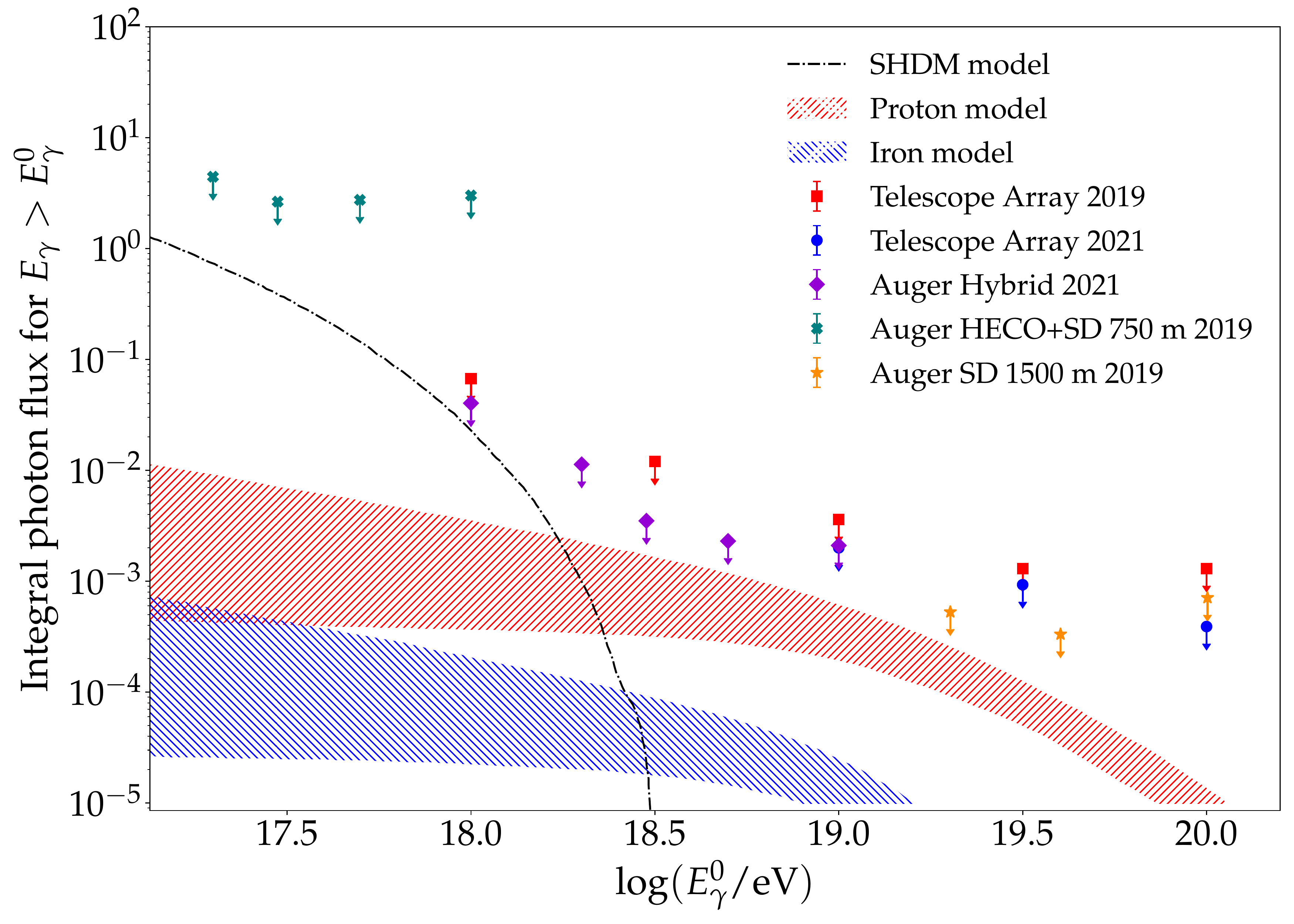}
\caption{\textls[-5]{Upper limits on the integral photon flux at $95\%$ confidence level obtained by 
Auger~\cite{AugerPh:19,AugerPh:21}} and Telescope Array~\cite{TAPh:19,TAPh:21}. The shaded regions correspond two models 
in which the ultra-high-energy cosmic rays are dominated by protons and iron nuclei~\cite{Kampert:12}. \textls[-15]{The 
dashed-dotted curve corresponds to the integral photon flux originated by the decay of hypothetical super-heavy dark 
matter particles of mass $M_\chi = 10^{10}$ GeV/c$^2$ and lifetime $\tau_\chi = 3 \times 10^{23}$ year~\cite{Kalashev:16}. 
Adapted from~\cite{AugerPh:21}}. \label{PhLimits}}
\end{figure}

High-energy photons arriving at Earth generate EASs that are almost electromagnetic. They are characterized by having
a much deeper $X_{\max}$ and very small muon content. For this reason, it is easier to separate hadrons from photons
than heavy from light hadrons. At present, there are only upper limits to the integrated photon flux obtained by 
different experiments. The most restrictive ones are obtained by Auger and Telescope Array, which are shown in 
Figure~\ref{PhLimits}. The figure also shows the integrated photon flux corresponding to two extreme models in which
the ultra-high-energy cosmic-ray flux is dominated by protons and by iron nuclei~\cite{Kampert:12}. From the figure, 
it can be seen that the upper limits are reaching the integrated flux values corresponding to the proton model.    

Photon searches at the highest energies have been motivated in the past by the top-down models, in which the ultra-high-energy cosmic rays originate in the decay of super-heavy relic particles present in the halo of our Galaxy or by the decay 
of topological defects (see Ref.~\cite{Batacharjee:00}). In these models, the ultra-high-energy cosmic-ray flux is dominated
by photons. These top-down models are disfavored by current data, but it is still possible that a minority component contributes
to the total flux. Figure~\ref{PhLimits} shows the high-energy integral photon flux due to the decay of super-heavy dark
matter particles, located in the halo of our Galaxy, of mass $M_\chi = 10^{10}$ GeV/c$^2$ (c is the speed of light) and 
lifetime \mbox{$\tau_\chi = 3 \times 10^{23}$ year~\cite{Kalashev:16}}. From the figure, it can be seen that such a model is still 
compatible with the most stringent upper limits found up to now.      

The neutrino search at the highest energies is conducted by using different experimental techniques. There are two main methods 
to search for neutrinos in cosmic-ray observatories. The first is based on the fact that the neutrino cross section is so 
small that it is more probable that a neutrino initiates a shower close to the horizontal direction and very deep in the 
atmosphere. One of these showers is very easy to identify, as hadronic or even electromagnetic showers develop 
significantly earlier in the atmosphere. The other method is based on the fact that tau neutrinos that propagate skimming
the Earth, of zenith angles larger than 90$^\circ$, can interact in the Earth and produce a tau lepton, which can decay in
the atmosphere, initiating a shower. These showers are also easy to identify because they develop in the up-going direction. 

There are also dedicated neutrino observatories, such as IceCube, which is a cubic-kilo\-me\-ter particle detector made 
of Antarctic ice. It is buried beneath the surface, reaching a depth of about 2.5 km. In this case, neutrinos are  
detected when they interact with the molecules of the ice, producing relativistic charged particles, which emit
Cherenkov radiation measured with optical modules. These optical modules are placed in a hexagonal grid of 85 
strings starting at 1450 m and reaching 2450 m depth. On the other hand, ANITA consists of an array of radio 
antennas installed on a balloon flying at a height of $\sim$$37$ km in Antarctica. ANITA was designed to measure 
impulsive radio emissions from a neutrinos-initiated cascade in the Antarctic ice. It can also observe extensive air 
showers induced by cosmic rays or other particles. 

\begin{figure}[ht!]
\centering
\includegraphics[width=12cm]{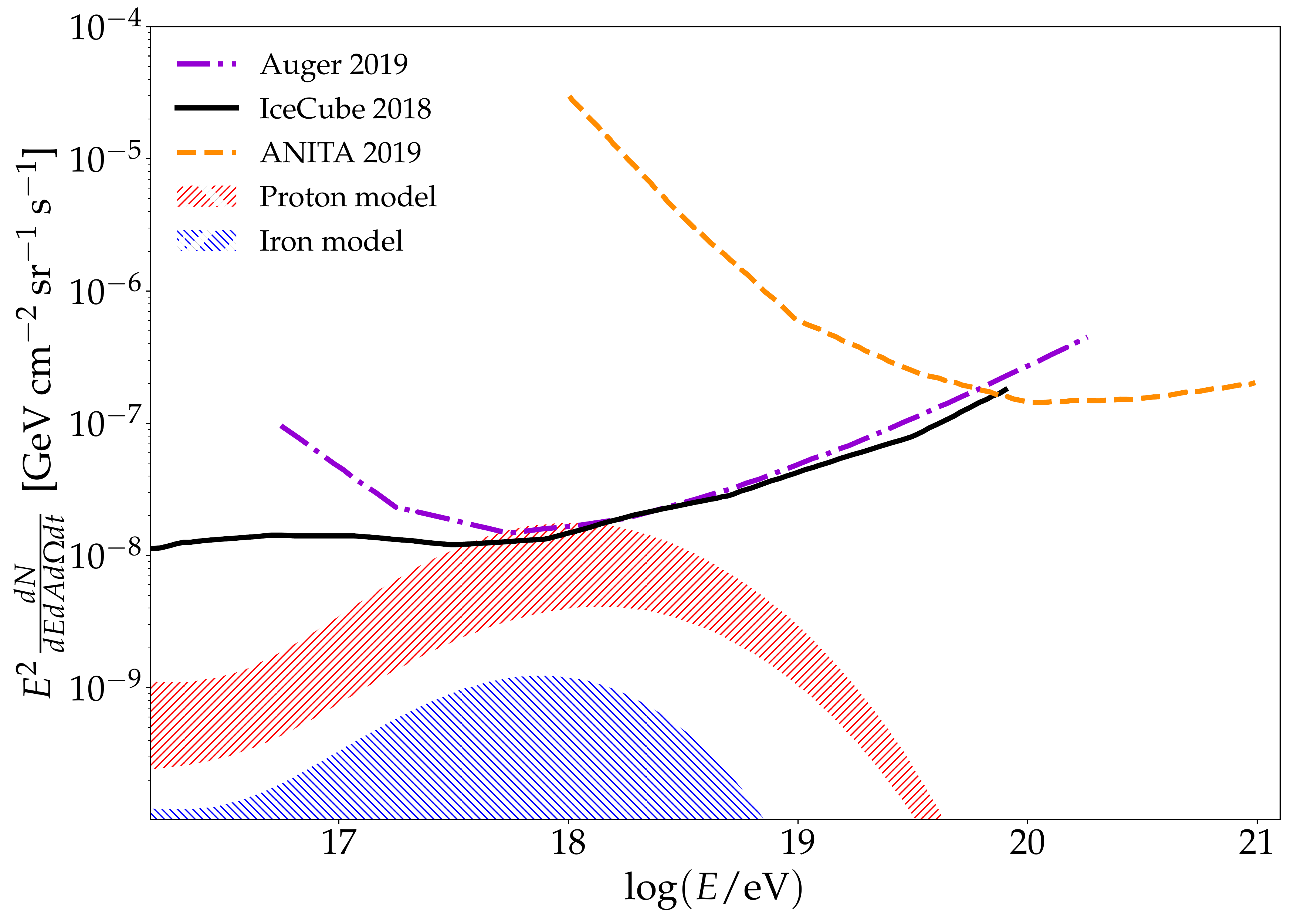}
\caption{Upper limits on the differential total neutrino flux at $90\%$ confidence level obtained by IceCube 
\cite{IceCubeUL:18}, Auger~\cite{AugerNuUL:10}, and ANITA~\cite{ANITA:19}. The shaded regions correspond to two models 
(same as the ones in Figure~\ref{PhLimits}) in which the ultra-high-energy cosmic rays are dominated by protons and iron
nuclei~\cite{Kampert:12}. 
\label{NuLimits}}
\end{figure}

Figure~\ref{NuLimits} shows the upper limits on the differential neutrino flux, at a $90\%$ confidence level, obtained by
IceCube~\cite{IceCubeUL:18}, Auger~\cite{AugerNuUL:10}, and ANITA~\cite{ANITA:19}, which are the most restrictive. In 
the figure, the expected neutrino flux for the same models considered before can also be seen. Note that the IceCube 
and Auger upper limits are reaching the flux level of the proton models.

As mentioned in the introduction, the composition at the highest energies is a crucial information for predicting the
ultra-high-energy photon and neutrino fluxes. Another important parameter is the maximum energies reached by the cosmic 
rays injected by the sources. The models that best fit the Auger flux and composition predict very low neutrino and 
photon fluxes~\cite{Batista:191}. However, the neutrino flux predicted by the models that best fit the 
Telescope Array flux and composition is larger~\cite{TAFit:21}. This is mainly due to the larger values of the fitted 
maximum energy and also, but to a lesser extent, to the fitted lighter composition. In any case, more statistics in the
suppression region are required to obtain a more accurate fit of the data.    

Additionally, the ultra-high-energy photons and neutrinos fluxes can constrain the composition of the ultra-high-energy cosmic 
rays independently of the high-energy hadronic interaction models~\cite{Hooper:11,Vliet:19}. As can be seen from Figures 
\ref{PhLimits} and \ref{NuLimits}, the neutrino and photon fluxes in models dominated by light nuclei are larger than
the ones dominated by heavy nuclei. This is mainly due to the fact that the threshold energy of the photo-pion production
increases with the mass number and, because the injection spectrum decreases with the cosmic ray energy, there are less 
particles in heavy nuclei dominated models that undergo photo-pion production.

\section{Future Perspectives on Composition}
\label{Future}

The data considered in previous sections were taken by detectors that are currently in operation. These experiments
will continue taking data and refining the methods used in composition analyses. An example of the progress in the 
methods used to study the composition is the increasing use of machine learning techniques 
\cite{IceCubeJ:19,TA:21,AugerDL:21,AugerRNN:21}, which have proven to be powerful tools for reconstructing composition 
sensitive parameters and also in the composition determination.  

Some of the current observatories will be or are being upgraded. In particular, Auger started an upgrade of the 
observatory, which has the detailed study of the composition from the transition region
up to the highest energies as one of its main objectives. The Auger upgrade is known as AugerPrime~\cite{AugerPrime:16,AugerPrime:18}. The improvements
relevant for composition are: the addition of plastic scintillation detectors on top of the water-Cherenkov detectors to 
separate the electromagnetic from the muonic components of the showers; the addition of underground muon detectors in the
infill region to measure the muonic component of the showers in the transition region; the extension of the current 
fluorescence telescopes measurements into periods of higher night-sky background in order to increase their duty cycle; and the addition of a radio detector to each water-Cherenkov station which can measure $X_{\textrm{max}}$ and other composition-sensitive parameters. These enhancements will increase the mass composition sensitivity of the observatory. One of the 
main goals of the upgrade is to measure the composition at the highest energies, in the suppression region, by using the 
upgraded surface detectors, which can collect about one order of magnitude more statistics than the fluorescence 
detectors. 

Additionally, Telescope Array started an upgrade of the observatory called TAx4~\cite{TAx4:21}. It consists of an  
increase in the detection area by adding new surface detectors and fluorescence telescopes. The detection area 
increased from $\sim$$700$ to $\sim$$2800$ km$^2$. With this new detection area, it will be possible to measure 
the composition from the northern hemisphere by using both surface and fluorescence detectors with much larger 
statistics.

IceCube is also planning an upgrade called the IceCube-Upgrade~\cite{IceCubeUp:21}. It will consist of an enhancement of
the IceTop surface detectors by adding scintillation detectors, radio detectors, and possibly small non-image Cherenkov 
telescopes. One of the main goals of this upgrade is also to increase the mass composition sensitivity of the observatory.     

Next generation observatories are currently being planned. In particular, IceCube-Gen2~\cite{IceCubeGen2}, the successor
of IceCube, will measure cosmic rays at low energies. It will consist of an optical array in the deep ice, a large-scale 
radio array, and a surface detector above the optical array. It will be able to measure composition with more sensitivity
and it will increase its maximum energy up to $\sim$$10^{18}$ eV\@. At the highest energies, the POEMMA project is designed
to measure ultra-high-energy cosmic rays and neutrinos from the space~\cite{POEMMA:21}. It will consist of two identical 
orbital fluorescence telescopes being able to measure the $X_{\textrm{max}}$ parameter at the highest energies with
unprecedented statistics. The GRAND project~\cite{GRAND:21} is designed to detect ultra-high-energy neutrinos, cosmic rays, 
and gamma rays by using the radio technique. It will consist of $2\times 10^5$ radio antennas covering an area of 
$2\times 10^5$ km$^2$, which will be separated into 20 sub-arrays of $\sim$$10^4$ km$^2$. It will be able to measure the
$X_{\textrm{max}}$ parameter in a wide energy range with large statistics. Finally, GCOS is a starting project intended 
to design a next-generation observatory with an aperture of at least one order of magnitude larger than the one 
corresponding to the observatories that are currently in operation~\cite{GCOS:21}. One of its main objectives is to 
improve the mass resolution, based on the experience obtained from AugerPrime and TAx4.

The data collected by the planned upgrades and also by the future observatories will constrain even more the high-energy
hadronic models used in shower simulations. Moreover, analyses like the one reported in Ref.~\cite{AugerMuFluct:21}, 
where the fluctuations of the $R_\mu$ parameter are studied and compared with model predictions, can contribute to 
understand more deeply the discrepancies found. Moreover, the data from the high-luminosity LHC~\cite{HLLHC:20} run will 
play an important role on the improvement of the current high-energy hadronic interaction models.

\section{Conclusions}
\label{Conc}

The origin of cosmic rays is still an open issue in high-energy astrophysics. The composition of the primary 
particle is key information to understanding this phenomenon. In the last years, a big effort has been made to determine 
it. The main limitation comes from the incompatibility of current high-energy hadronic interaction models, used to 
simulate the EAS required to determine the primary mass, with the experimental data. 

The determination of the composition is mainly achieved by using optical detectors that measure the depth of the shower 
maximum. In particular, the mean value of $X_{\textrm{max}}$ measured by different experiments falls between the proton 
and iron expectations obtained by using current high-energy hadronic interaction models. There are also several composition 
parameters that come from the surface detector of which the most sensitive to the nature of the primary is the muon content 
of the showers. In general, the composition obtained from parameters measured by surface detectors is incompatible with the 
one obtained from $X_{\textrm{max}}$. These discrepancies are larger at the highest energies, where the composition obtained 
by some of the parameters falls above the expectations for iron nuclei, obtained by using current models. These
incompatibilities are usually interpreted as a deficit in the number of muons predicted by current high-energy hadronic
interaction models. Even though the hadronic part of the simulated cascades seems to be the source of the discrepancies, 
the electromagnetic part, which affects the $X_{\textrm{max}}$ predictions, can also play an important role as showed by
recent analyses.

In the energy interval between $10^{15}$ and $10^{18}$ eV, the composition obtained from the $\langle X_{\textrm{max}} \rangle$ 
measured by different experiments presents large differences. The composition measurement in this energy range is very
important for understanding the nature of the second knee, which can be interpreted as the region where the transition 
between the galactic and extragalactic component takes place, as suggested by the KASCADE-Grande data. The IceCube data cannot
corroborate this hypothesis, but this can be due to the systematic uncertainties introduced by current high-energy hadronic 
interaction models used to analyze the experimental data, which were showed to be more important in composition analyses based 
on surface detectors information.  

Above $\sim$$10^{18}$ eV, the composition measurements obtained by different experiments that measured the
$\langle X_{\textrm{max}} \rangle$ are compatible within statistical and systematic uncertainties. The composition seems
to be light between $10^{18}$ and $10^{18.5}$ eV\@. At higher energies, the composition is compatible with an increasingly 
heavier average mass as a function of primary energy. In this energy region, the discrepancies between the composition 
obtained from the $X_{\textrm{max}}$ parameter and the muon component of the showers are the largest. 

The importance of the composition information on many aspects of cosmic ray studies has become more relevant in recent
years. This motivated several upgrades of current observatories with a special interest in the increase in the mass sensitivity 
of the detectors. Moreover, the next generation of cosmic-ray observatories are also being designed, taking into account the 
importance of the mass composition determination. Moreover, the planned, much larger detection areas in combination with the enhanced 
mass composition sensitivity of future detectors will allow to constrain or even measure the photon and neutrino fluxes at the 
highest energies, which can be used to constrain the composition in the suppression region.

An improvement on current high-energy hadronic interaction models is required to make progress on the composition 
determination. This can be achieved in the near future, based on the synergy between the EAS studies and the ones conducted 
in particle physics, especially with the future high-luminosity LHC data.

\vspace{6pt} 




\funding{This research was supported by CONICET, PIP 11220200102979CO.}

\institutionalreview{Not applicable.}

\informedconsent{Not applicable.}

\acknowledgments{A.D.S. is member of the Carrera del Investigador Cient\'iﬁco of CONICET, Argentina. The author thanks Armando di Matteo for a careful reading of the manuscript and to the members of the Pierre Auger Collaboration for useful
discussions.}

\conflictsofinterest{The authors declare no conflict of interest.}


\begin{adjustwidth}{-\extralength}{0cm}
\reftitle{References}

\end{adjustwidth}
%


\end{document}